\documentclass[ prb,twocolumn, showkeys,eqsecnum,amsmath,amssymb]{revtex4}
\usepackage{graphicx}
\usepackage{dcolumn}
\usepackage{bm}
\begin{document}
\title{
Strong-Coupling Superconductivity in the Cuprate Oxide
}
\author{Fusayoshi J. O{\scriptsize HKAWA}%
\footnote{E-mail: fohkawa@mail.sci.hokudai.ac.jp}}
\affiliation{Department of Physics, Faculty of Science, 
Hokkaido University, Sapporo 060-0810, Japan}
\received{May 17, 2009; accepted May 26, 2009; published August 10, 2009%
\footnote{Published in J. Phys. Soc. Jpn. {\bf 78} (2009) 084712 (1-17).\\
\phantom{$\dagger$}DOI: 10.1143/JPSJ.78.084712}}
%
\begin{abstract}
Superconductivity in the cuprate oxide is studied by Kondo-lattice theory based on the $t$-$J$ model with the electron-phonon interaction arising from the modulation of the superexchange interaction by phonons. 
The self-energy of electrons is decomposed into the single-site and multisite self-energies. 
It is proved by using the mapping of the single-site self-energy in the $t$-$J$ model to its corresponding one in the Anderson model that 
the single-site self-energy is simply that of a conventional Fermi liquid, even if a superconducting order parameter appears or the multisite self-energy is anomalous. 
The electron liquid characterized by the single-site self-energy
is a conventional Fermi liquid. The Fermi liquid is further stabilized by the resonating-valence-bond (RVB) mechanism. The stabilized Fermi liquid is a relevant {\it unperturbed} state that can be used to study superconductivity and anomalous Fermi-liquid behaviors. 
The so-called spin-fluctuation-mediated exchange interaction, which includes the superexchange interaction as a part, is the attractive interaction that binds $d_{x^2-y^2}$-wave Cooper pairs.
An analysis of the spin susceptibility implies that, because of 
the electron-phonon interaction, the imaginary part of the exchange interaction has a sharp peak or dip at $\pm\omega^*$, where $\omega^*\simeq \omega_{\rm ph}$ in the normal state and $\frac1{2}\epsilon_{\rm G} \lesssim \omega^*\lesssim \frac1{2}\epsilon_{\rm G} + \omega_{\rm ph}$ in the superconducting state, where $\omega_{\rm ph}$ is the energy of relevant phonons and $\epsilon_{\rm G}$ is the superconducting gap.
If the imaginary part has a sharp peak or dip at $\pm\omega^*$, then the dispersion relation of quasi-particles has kink structures near $\pm\omega^*$ above and below the chemical potential, the density of states has dip-and-hump structures near $\pm \omega^*$ outside the coherence peaks in the superconducting state, and the anisotropy of the gap deviates from the simple $d_{x^2-y^2}$-wave anisotropy. \\
\end{abstract}
%
%
%
 %
%
\maketitle
\newcommand\mysection[1]{\section{\NoCaseChange{#1}}}
\newcommand\mysubsection[1]{\subsection{\it \NoCaseChange{#1}}}
\newcommand\mysubsubsection[1]{\subsubsection{\NoCaseChange{#1}}}
\renewcommand\thesection{\arabic{section}}
\renewcommand\thesubsection{\arabic{section}.\arabic{subsection}}
\renewcommand\thesubsubsection{\arabic{section}.\arabic{subsection}.\arabic{subsubsection}}
\mysection{Introduction}
\label{SecIntroduction}
%
%
Although many experimental and theoretical studies have been performed since the discovery of the cuprate-oxide superconductor in 1986, \cite{bednortz,RevD-wave,RevScience,RevStripe,lee,RevScanning} the mechanism for high-temperature (high-$T_c$) superconductivity in the cuprate oxide is still contentious.
Recently, Fe-based superconductors have been discovered.
\cite{hosono,day}
Since both cuprate-oxide and Fe-based superconductors are in the vicinity of the Mott metal-insulator transition, it is certain that strong electron correlations play a crucial role in the mechanism for high-$T_c$ superconductivity. 
The most crucial issue is whether the {\it normal} state is actually an exotic Fermi liquid such as the resonating-valence-bond (RVB) state \cite{RVB1987}
or it is nevertheless a conventional Fermi liquid.


One of the simplest effective Hamiltonians for an electron liquid in the vicinity of the Mott transition is the Hubbard model. According to Hubbard's theory, \cite{Hubbard1,Hubbard2} when the on-site repulsion $U$ is so large that $U\gtrsim W$, where $W$ is the bandwidth, the Hubbard gap opens between the upper and lower Hubbard bands. 
According to Gutzwiller's theory, \cite{Gutzwiller1,Gutzwiller2,Gutzwiller3} together with the Fermi-liquid theory, \cite{Luttinger1,Luttinger2} a narrow quasi-particle band appears near the chemical potential; this band is called the Gutzwiller band in this paper.
One may speculate that the density of states has a three-peak structure, with the Gutzwiller band between the upper and lower Hubbard bands.  The approximations used in
Hubbard's and Gutzwiller's theories are called Hubbard and Gutzwiller approximations, respectively, both of which are within the single-site approximation (SSA). According to another SSA theory, \cite{OhkawaSlave} the Gutzwiller band appears at the top of the lower Hubbard band for $n<1$, where $n$ is the electron density per unit cell, which implies that it appears at the bottom of the upper Hubbard band for $n>1$. 
The SSA that considers all the single-site terms is rigorous for $d\rightarrow +\infty$ within the restricted Hilbert subspace where no order parameter exists, \cite{Metzner,Muller-H1,Muller-H2,Janis} 
where $d$ is the spatial dimensionality; this SSA is called the supreme single-site approximation (S$^3$A) in this paper.
The S$^3$A theory is reduced or mapped to a problem of self-consistently determining and solving the Anderson model, \cite{Mapping-1,Mapping-2,Mapping-3,georges} which is an effective Hamiltonian for studying the Kondo effect. 
The three-peak structure corresponds to that in the Anderson model, with the Kondo peak between two subpeaks. 
The Kondo effect has relevance to electron correlations in the vicinity of the Mott transition.
The S$^3$A is also formulated as the dynamical mean-field theory\cite{georges,RevMod,kotliar,PhyToday} (DMFT) and the dynamical coherent potential approximation \cite{dpca} (DCPA).
These three formulations are exactly equivalent to each other.

According to Brinkman and Rice's theory,\cite{brinkman} which also uses the Gutzwiller approximation, the Gutzwiller band vanishes when $n=1$ and $U> U_\text{BR}$, with $U_\text{BR}\simeq W$.
Cluster DMFT (CDMFT) \cite{cellDMFT1,cellDMFT2,cellDMFT3,cellDMFT4} 
has been proposed as an extension of DMFT. 
According to the numerical results of DMFT
\cite{RevMod,kotliar,PhyToday}
and CDMFT, \cite{cellDMFT1,cellDMFT2,cellDMFT3,cellDMFT4} 
when $n\simeq 1$ and $U\gtrsim U_\text{BR}$ the Gutzwiller band appears to vanish and the ground state within the restricted Hilbert subspace appears to be a Mott insulator. The numerical DMFT and CDMFT are consistent with Brinkman and Rice's theory, which treats the case of $n=1$, but are inconsistent with Gutzwiller's theory for $n\ne 1$, which predicts that the ground state is a metal for $n\ne 1$.

In a previous paper,\cite{proofFL} it has been proved that the ground state in the S$^3$A is a conventional Fermi liquid except when $n=1$ and $U/W=+\infty$, at least if an electron reservoir is {\it explicitly} considered in the grand canonical ensemble.
It is surprising that the  ground states are different between DMFT and the S$^3$A, which are equivalent to each other.
This discrepancy is presumably because the electron reservoir is not explicitly considered in the numerical DMFT, while it is explicitly considered in the proof. \cite{comment1}
%
%
According to the proof, the conventional Fermi liquid constructed in the S$^3$A is a relevant {\it unperturbed} state that can be used to study anomalous Fermi-liquid behaviors and ordered states in the vicinity of the Mott transition.
A perturbative theory starting from the unperturbed state
is simply Kondo-lattice theory, \cite{Mapping-1,Mapping-2,Mapping-3} 
in which local electron correlations are, in principle, {\it exactly} treated in the S$^3$A, intersite electron correlations are treated by a conventional perturbation, and the spontaneous appearance of an order parameter is treated by an anomalous perturbation.

Although a multiband model is needed to explain the overall and precise features of both the cuprate-oxide and Fe-based superconductors, a single-band model can be approximately used, at least for the cuprate oxide. 
For example, an effective Hamiltonian for the cuprate oxide is the $d$-$p$ model, which considers $d$ orbits on Cu ions and $p$ orbits on O ions on CuO$_2$ planes.
When the on-site $U$ is sufficiently strong, the $d$-$p$ model is approximately mapped to the $t$-$J$ model, \cite{rice} which is a single-band model.
The $t$-$J$ model is also derived from the Hubbard model. \cite{hirsch}
The $t$-$J$ model is one of the simplest effective Hamiltonians for an electron liquid in the vicinity of the Mott transition or in the cuprate oxide.

A mechanism for high-$T_c$ superconductivity in the cuprate oxide was proposed in 1987: \cite{FJO1987} the condensation of $d_{x^2-y^2}$-wave Cooper pairs between heavy quasi-particles by the superexchange interaction.
Since the superexchange interaction is as strong as $J=-(0.10\mbox{-}0.15)$~eV between nearest neighbors,\cite{SuperJ} it is certain to play a role in the binding of $d_{x^2-y^2}$-wave Cooper pairs in the cuprate oxide.
On the other hand, there is evidence that the electron-phonon interaction is strong in the cuprate oxide:
the softening of the half-breathing modes
near $\left(\pm 1, 0\right)(\pi/a)$ and $\left(0, \pm 1\right)(\pi/a)$
in the two-dimensional Brillouin zone,
\cite{McQ1,Pint1,McQ2,Pint2,Braden}
where $a$ is the lattice constant of the CuO$_2$ planes,
the softening of Cu-O bond stretching modes
near $\left(\pm 1/2,0\right)(\pi/a)$ and $\left(0,\pm 1/2\right)(\pi/a)$, 
\cite{pintschovius,reznik} 
and kinks in the dispersion relation of quasi-particles. 
\cite{lanzara-soft,johnson,tsato} 
A dip-and-hump structure outside the coherence peaks in the density of states has also been observed in the superconducting state by tunnelling spectroscopy.\cite{dip-hump} A similar structure is observed in  a conventional Bardeen-Cooper-Schrieffer (BCS) superconductor and is regarded as evidence that the binding of Cooper pairs is mainly due to the conventional electron-phonon interaction.\cite{mcmillan,scalapino,carbotte}
However, the conventional electron-phonon interaction arises from the charge-channel interaction; thus, it can never be strong in the vicinity of the Mott transition because charge fluctuations are suppressed there. If the electron-phonon interaction is strong in the cuprate oxide, it must arise from the spin-channel interaction such as that arising from the modulation of the superexchange interaction by phonons.\cite{FJO-elph1,FJO-elph2}
It will be interesting to study the role that the strong electron-phonon interaction plays in the mechanism for high-$T_c$ superconductivity. 

One of the purposes of this paper is to formulate a theory of superconductivity based on the $t$-$J$ model with the electron-phonon interaction. The other purpose is to apply it to superconductivity in the cuprate oxide.
The preliminaries are given in \mbox{\S\hskip2pt\ref{SecPreliminaries}}.
The Kondo-lattice theory of superconductivity is formulated in \mbox{\S\hskip2pt\ref{SecFormulation}}.
In \mbox{\S\hskip2pt\ref{SecApplication}}, the formulation is applied to strong-coupling superconductivity in the cuprate oxide.
A discussion is given in \mbox{\S\hskip2pt\ref{SecDiscussion}} and
a conclusion is given in \mbox{\S\hskip2pt\ref{SecConclusion}}.
An inequality that is crucial in the formulation in \mbox{\S\hskip2pt\ref{SecFormulation}} is proved in Appendix~\ref{SecProof}. 
The dynamical spin susceptibility of the $t$-$J$ model is studied in Appendix~\ref{SecDynamicalSus}.
The possibility of the spontaneous appearance of antiferromagnetic moments in the superconducting state is examined in Appendix~\ref{SecSG-AM}.

\mysection{Preliminaries}
\label{SecPreliminaries}
\mysubsection{Electron-phonon interaction}
\label{EP-interaction}
In the $d$-$p$ model, according to field theory, the superexchange interaction arises from the virtual exchange of a pair excitation of $d$ electrons between the upper and lower Hubbard bands.\cite{OhSupJ1}
When nonzero bandwidths of the Hubbard bands are ignored, the superexchange interaction constant is given by
%
\begin{equation}\label{EqSuperJ-1}
J =-  \frac{4V^4}{(\epsilon_d-\epsilon_p+U)^2}
\left(\frac{1}{\epsilon_d-\epsilon_p+U}+\frac{1}{U} \right) ,
\end{equation}
%
between $d$ electrons on nearest-neighbor Cu ions, where $V$ is the hybridization matrix between nearest neighbor $p$ and $d$ orbits, $\epsilon_d$ and $\epsilon_p$ are the depths of $d$ and $p$ orbits, respectively, and $U$ is the on-site repulsion between $d$ electrons.
Equation~(\ref{EqSuperJ-1}) is exactly the same as that derived by the fourth-order perturbation in $V$.
When nonzero bandwidths of the Hubbard bands are considered, $J$ is smaller than the value given by eq.~(\ref{EqSuperJ-1}).

We assume that the Hamiltonian for phonons in the cuprate oxide is given by
${\cal H}_{\rm ph} = \sum_{\lambda{\bm q}}\omega_{\lambda{\bm q}} \bigl(b_{\lambda{\bm q}}^\dag b_{\lambda{\bm q}} +1/2\bigr)$,
where $b_{\lambda{\bm q}}^\dag$ and $b_{\lambda{\bm q}}$
are the creation and annihilation operators, respectively, of a phonon with 
mode $\lambda$ and wave-number vector ${\bm q}$, and
$\omega_{\lambda{\bm q}}$ 
is the energy of the phonon.
The displacements of the $i$th Cu ion and the $[ij]$th O ion,
which lies between the nearest-neighbor $i$th and $j$th Cu ions, are given by
%
\begin{equation}\label{EqDispCu}
{\bm u}_i = 
 \sum_{\lambda{\bm q}}
 \frac{\hbar v_{d,\lambda{\bm q}} } 
{\sqrt{ 2N M_d \omega_{\lambda{\bm q}}} } 
e^{i{\bm q}\cdot{\bm R}_i}
{\bm e}_{\lambda{\bm q}} \left(
b_{\lambda{\bm q}}^\dag + b_{\lambda-{\bm q}} \right), 
\end{equation}
and
%
\begin{equation}\label{EqDispO}
\hskip-1pt {\bm u}_{[ij]} =
\sum_{\lambda{\bm q}}
\frac{\hbar v_{p,\lambda{\bm q}}}
{\sqrt{2N M_p \omega_{\lambda{\bm q}}} } 
 e^{i{\bm q}\cdot {\bm R}_{[ij]} }
{\bm e}_{\lambda{\bm q}} \left(
b_{\lambda{\bm q}}^\dag + b_{\lambda-{\bm q}} \right), 
\end{equation}
where $v_{d,\lambda{\bm q}}$ and $v_{d,\lambda{\bm q}}$ are real, ${\bm R}_i$ and ${\bm R}_{[ij]} = ({\bm R}_i + {\bm R}_j)/2$ are the positions of the $i$th Cu and $[ij]$th O ions, $M_d$ and $M_p$ are the masses of Cu and O ions, respectively, $N$ is the number of unit cells, 
and ${\bm e}_{\lambda{\bm q}}=(e_{\lambda {\bm q},x},e_{\lambda {\bm q},y},e_{\lambda {\bm q},z})$ is the polarization vector. 
%


It is convenient to define a {\it dual-spin} operator by
\begin{equation}\label{EqTwoSpin1}
\! {\cal P}_\Gamma({\bm q}) = \frac1{2} 
\sum_{{\bm q}^\prime}\eta_{1\Gamma}({\bm q}^\prime)  \left[
{\bm S}\left({\bm q}^\prime \!+\! \mbox{$\frac{1}{2}$}{\bm q}\right)
\!\cdot \!{\bm S}\left(-{\bm q}^\prime \!+\! \mbox{$\frac{1}{2}$}{\bm q}
\right)\right] ,
\end{equation}
where
\begin{equation}
{\bm S}({\bm q})= \frac1{\sqrt{N}} \sum_{\bm k\sigma\sigma^\prime} 
\frac1{2}{\bm \sigma}^{\sigma\sigma^\prime} d_{({\bm k} +\frac{1}{2}{\bm q})
\sigma}^\dag d_{({\bm k} -\frac{1}{2}{\bm q}) \sigma^\prime}. 
\end{equation}
Here, ${\bm \sigma} = \left(\sigma_x , \sigma_y ,\sigma_z \right)$
is the Pauli matrix, 
$d_{{\bm k}\sigma}^\dag$ and $d_{{\bm k}\sigma}$
are the creation and annihilation operators of $d$ electrons, respectively, and $\eta_{1\Gamma}({\bm q})$ are form factors defined by
\begin{equation}\label{EqForm1S}
\eta_{1s}({\bm q}) = \cos(q_xa) + \cos(q_ya),
\end{equation}
and
\begin{equation}\label{EqForm1D}
\eta_{1d}({\bm q}) = \cos(q_xa) - \cos(q_ya) .
\end{equation}
%
It is assumed that the $x$ and $y$ axes are within CuO$_2$ planes and that the $z$ axis is perpendicular to the CuO$_2$ planes.
Two types of electron-phonon interaction arise from the modulation of $J$ by the vibrations of O and Cu ions,\cite{FJO-elph1,FJO-elph2} which are denoted by ${\cal H}_p$ and ${\cal H}_d$, respectively: 
\begin{subequations}\label{EqEl-Ph}
\begin{align}
{\cal H}_p &=
C_p \sum_{\lambda{\bm q} }
\frac{\hbar v_{p,\lambda{\bm q}}}
{\sqrt{2 N M_p \omega_{\lambda{\bm q}}}} 
\left(b_{\lambda{\bm q}}^\dag + b_{\lambda-{\bm q}} \right)
\nonumber \\ &  \quad \times
\bar{\eta}_{\lambda s}({\bm q}) \sum_{\Gamma=s,d} 
\eta_{1\Gamma}\left(\mbox{$\frac{1}{2}{\bm q}$}\right) 
{\cal P}_\Gamma({\bm q}) , 
\end{align}
and
\begin{align}
{\cal H}_d &=
C_d \sum_{\lambda{\bm q}} 
\frac{\hbar v_{d,\lambda{\bm q}}}
{\sqrt{2 N M_d \omega_{\lambda{\bm q}}}} 
\left(b_{\lambda{\bm q}}^\dag + b_{\lambda-{\bm q}} \right)
\nonumber \\ &  \quad \times
\sum_{\Gamma=s,d} 
\bar{\eta}_{\lambda \Gamma}({\bm q})
{\cal P}_\Gamma({\bm q}) , 
\end{align}
\end{subequations}
where $C_p$ and $C_d$ are real constants, which are given in a previous paper,\cite{FJO-elph1}
\begin{equation}
\bar{\eta}_{\lambda s}({\bm q}) =
2\left[ e_{\lambda {\bm q}, x} \sin\left(\frac{q_x a}{2}\right) 
+  e_{\lambda {\bm q},y} \sin\left(\frac{q_y a}{2}\right)\right] ,
\end{equation}
and
\begin{equation}
\bar{\eta}_{\lambda d}({\bm q}) =
2\left[ e_{\lambda {\bm q}, x} \sin\left(\frac{q_x a}{2}\right) 
-  e_{\lambda{\bm q}, y} \sin\left(\frac{q_y a}{2}\right)\right] .
\end{equation}

Since the antiferromagnetic $J$ is present between nearest neighbors, spin fluctuations develop near ${\bm Q}_M=(\pm 1, \pm 1)(\pi/a)$.
\cite{comSquareLattice}
We denote the center wave number of spin fluctuations by
${\bm Q}_{\rm sf}= {\bm Q}_M+\Delta {\bm Q}_{\rm sf}$, where $|\Delta {\bm Q}_{\rm sf}|a\ll 1$.
According to eqs.~(\ref{EqTwoSpin1}) and (\ref{EqEl-Ph}),
two modes of spin fluctuations with wave number ${\bm Q}_{\rm sf}$
couple with a mode of phonons with wave number ${\bm Q}_{\rm ph} = 2{\bm Q}_{\rm sf}\pm{\bm G}$ or ${\bm Q}_{\rm ph} = 2\Delta {\bm Q}_{\rm sf}$, where ${\bm G}$ is a reciprocal lattice vector. 
When spin fluctuations develop near ${\bm Q}_M+\Delta {\bm Q}_{\rm sf}$, therefore, phonons can be soft near $2\Delta{\bm Q}_{\rm sf}$. \cite{FJO-elph2}
The electron-phonon interaction vanishes in the limit of $|{\bm Q}_{\rm ph}|a\rightarrow 0$ or $|\Delta{\bm Q}_{\rm sf}|a\rightarrow 0$ but it is nonzero for nonzero ${\bm Q}_{\rm ph}$ or nonzero $\Delta{\bm Q}_{\rm sf}$.
When the electron-phonon interaction is strong, ${\bm Q}_{\rm sf}$ cannot be exactly ${\bm Q}_M$ although ${\bm Q}_{\rm sf}\simeq {\bm Q}_M$;
$\Delta {\bm Q}_{\rm sf}\ne 0$ but $\Delta {\bm Q}_{\rm sf}\simeq 0$.

When coupled modes between spin fluctuations and phonons are sharp, 
\begin{align}\label{EqTwoSpin2}
{\cal P}_{s}({\bm q})&=
{\cal P}_{s}({\bm q}\pm {\bm G}) = \sum_{{\bm q}^\prime}  
{\bm S}\left({\bm q}^\prime + \mbox{$\frac{1}{2}$}
{\bm q}\right)
\cdot {\bm S}\left(-{\bm q}^\prime  + \mbox{$\frac{1}{2}$}
{\bm q}\right) ,
\end{align}
and ${\cal P}_{d}({\bm q})= 0$ 
can be used as an approximation instead of eq.~(\ref{EqTwoSpin1}) because $\eta_{1s}(0)=2$ and $\eta_{1d}(0)=0$. The total electron-phonon interaction,
${\cal H}_\text{el-ph} ={\cal H}_p+{\cal H}_d$, is given by
\begin{align}
{\cal H}_\text{\rm el-ph} &=
\frac1{\sqrt{N}}\sum_{\lambda{\bm q} }K_\lambda({\bm q})
\left(b_{\lambda{\bm q}}^\dag + b_{\lambda-{\bm q}} \right)
{\cal P}_{s}({\bm q}) , 
\end{align}
where ${\cal P}_{s}({\bm q})$ is given by eq.~(\ref{EqTwoSpin2}) and
\begin{align}\label{EqCouplSfPh}
K_\lambda({\bm q}) &=
\frac{\hbar}{\sqrt{2\omega_{\lambda{\bm q} }}}
\bar{\eta}_{\lambda s}({\bm q})
\left[\frac{C_p v_{p,\lambda{\bm q}}}{\sqrt{M_p}}
\eta_{1s}\hskip-2pt\left(\mbox{$\frac{1}{2}{\bm q}$}\right)
+\frac{C_d v_{d,\lambda{\bm q}}}{\sqrt{M_d}}\right].
\end{align}

\mysubsection{Total effective Hamiltonian}
%
The $t$-$J$ or $t$-$J$-$U_\infty$ model is defined by
\begin{align}\label{Eqt-J-U}
{\cal H}_{t\text{-}J} &= 
\epsilon_a \sum_{i\sigma}n_{i\sigma} +
\sum_{i\ne j\sigma} t_{ij}
d_{i\sigma}^\dag d_{j\sigma}
- \frac1{2} J \sum_{\left<ij\right>}\left({\bm S}_i\cdot{\bm S}_j\right)
\nonumber \\ & \quad
+U_\infty \sum_{i} n_{i\uparrow} n_{i\downarrow},
\end{align}
where $\epsilon_a$ is the band center, $n_{i\sigma}=d_{i\sigma}^\dag d_{i\sigma}$, $t_{ij}$ are transfer integrals, the summation $\left<ij\right>$ runs over pairs of nearest-neighbor unit cells, and
${\bm S}_i = (1/2)\sum_{\sigma\sigma^\prime}
{\bm \sigma}^{\sigma\sigma^\prime}
d_{i\sigma}^\dag d_{i\sigma^\prime}$.
%
The dispersion relation of bare electrons is given by 
\begin{equation}\label{EqBareDispersion}
E({\bm k}) = 
\epsilon_a + 
\frac1{N} \sum_{i\ne j} t_{ij} \exp\left[{\rm i}{\bm k}\cdot
\left({\bm R}_i-{\bm R}_j\right)\right] ,
\end{equation}
where ${\bm R}_i$ is the position of the $i\hskip1pt$th unit cell. 
The bandwidth of $E({\bm k})$ is denoted by $W$. 
The on-site repulsion $U_\infty$ must be infinitely large to exclude any double occupancy. For the sake of convenience, $U_\infty$ is first treated as being finite and then the limit of 
%
$U_\infty/W\to+\infty$
%
is taken in final results. 
It is assumed in this paper that the electron density per unit cell, $n= (1/N)\sum_{i\sigma}\left<n_{i\sigma}\right> $, 
is less than that at half filling such that $0< n <1$.\cite{comFilling}

Following a previous paper, \cite{proofFL} we explicitly consider an electron reservoir in the grand canonical ensemble. 
The reservoir is defined by
%
${\cal H}_{\rm res} = \sum_{ij\sigma}
t_{ij}^\prime b_{i\sigma}^\dag b_{j\sigma}$,
%
and an infinitesimally small but random {\it hybridization} between the reservoir and the $t$-$J$-$U_\infty$ model is defined by
\begin{equation}\label{EqResV}
{\cal V} = {\lambda}_{0} \sum_{(ij) \in {\cal R}}\left[
 v_{(ij)} d_{i\sigma}^\dag b_{i\sigma} 
+ v_{(ij)}^* b_{i\sigma}^\dag d_{i \sigma} \right] ,
\end{equation}
where $\lambda_{0}=\pm 0^+$ is a nonzero but infinitesimally small numerical constant. In eq.~(\ref{EqResV}),
the summation $(ij)$ runs over pairs of sites, the $i$th site in the $t$-$J$-$U_\infty$ model and the $j$th site in the reservoir, in the set  ${\cal R}$.
Here,  %
it is assumed that
$\bigl<\hskip-3pt\bigl<\hskip1pt v_{(ij)}
\hskip1pt\bigr>\hskip-3pt\bigr> =
\bigl<\hskip-3pt\bigl<\hskip1pt  v_{(ij)}^*
\hskip1pt\bigr>\hskip-3pt\bigr> =0$ 
and 
\begin{equation}
\bigl<\hskip-3pt\bigl<\hskip1pt v_{(ij)} v_{(i^\prime j^\prime)}^*
\hskip1pt\bigr>\hskip-3pt\bigr> =
\delta_{(ij)(i^\prime j^\prime)} n_{\rm h}|v|^2, 
\end{equation}
where $\left<\hskip-2pt\left<\hskip1pt\cdots\hskip1pt\right>\hskip-2pt\right>$ denotes the ensemble average for ${\cal R}$ and $n_{\rm h}$ is the density of hybridization sites per unit cell, referring to the lattice on which the $t$-$J$-$U_\infty$ model exists.

The total Hamiltonian to be considered is given by
\begin{equation}
{\cal H} = {\cal H}_{t\mbox{-}J} +  
{\cal H}_{\text{el-ph}}+ {\cal H}_{\rm ph}
+ {\cal H}_{\rm res} + {\cal V}-\mu {\cal N}_{\rm t},
\end{equation}
where $\mu$ is the chemical potential and
${\cal N}_{\rm t} = \sum_{i\sigma} d_{i\sigma}^\dag d_{i\sigma}
+ \sum_{i\sigma} b_{i\sigma}^\dag b_{i\sigma}$.
%
For the sake of simplicity, it is assumed that the inversion symmetry exists in the total system averaged over the ensemble.

When $U_\infty=0$, $J=0$, and the electron-phonon interaction is absent, the single-particle Green function for $d$ electrons in the $t$-$J$-$U_\infty$ model averaged over the ensemble is given by
\begin{equation}\label{EqG0}
G_{\sigma}^{(0)}({\rm i}\varepsilon_n,{\bm k})=
\frac1{{\rm i}\varepsilon_n +\mu - E({\bm k}) 
- \Gamma ({\rm i}\varepsilon_n) } .
\end{equation}
Here, $\Gamma ({\rm i}\varepsilon_n)$ is the self-energy due to the hybridization with the reservoir. 
Since $\lambda_{0}=\pm 0^+$, the second-order perturbation is sufficiently accurate to treat $\Gamma ({\rm i}\varepsilon_n)$ as 
%
\begin{equation}\label{EqSigmaL}
\Gamma ({\rm i}\varepsilon_n) = 
n_{\rm h}\lambda_{0}^2 |v|^2\frac1{N_b}\sum_{\bm k}
\frac1{{\rm i}\varepsilon_n +\mu -E_b({\bm k})},
\end{equation}
%
where $N_b$ is the number of unit cells in the reservoir and
$E_b({\bm k}) = (1/N_b)\sum_{ij}t_{ij}^\prime
\exp[{\rm i}{\bm k}\cdot({\bm R}_i^\prime -{\bm R}_j^\prime)]$,
where ${\bm R}_i^\prime$ is the position of the $i\hskip1pt$th unit cell in the reservoir.
It is assumed that no gap opens in the reservoir at the chemical potential:
%
\begin{equation}\label{EqGamma}
-\mbox{Im} \hskip2pt \Gamma (+{\rm i}0)=0^+ > 0.
\end{equation}
%
The electron number within the $t$-$J$-$U_\infty$ model is never a constant of motion in the presence of the electron reservoir; thus, in general, the quantum-mechanically averaged number of electrons is a non-integer or an irrational number.
The presence of the reservoir with $N_b\rightarrow+\infty$ or eq.~(\ref{EqGamma}) ensures the quality of the grand canonical ensemble that the averaged number of electrons is a continuous function of the chemical potential.

\mysubsection{Fermi-surface condition}
\label{SecFScondition}
%
The Anderson model is defined by
\begin{align}\label{EqAnderson}
{\cal H}_{\rm A} &=
\sum_{{\bm k}\sigma} E_c({\bm k}) 
c_{{\bm k}\sigma}^\dag c_{{\bm k}\sigma}
+ \epsilon_d \sum_{\sigma}n_{d\sigma}
+ \tilde{U}_\infty n_{d\uparrow} n_{d\downarrow}
\nonumber \\ & 
+ \frac1{\sqrt{ N_{c} }} \sum_{{\bm k}\sigma} \left(
V_{\bm k}c_{{\bm k}\sigma}^\dag d_\sigma
+ V_{\bm k}^* d_\sigma^\dag c_{{\bm k}\sigma}\right) ,
\end{align} 
%
where $n_{d\sigma}=d_{\sigma}^\dag d_{\sigma}$ and $N_{c}$ is the number of unit cells. 
The Green function for $d$ electrons is given by
\begin{equation}
\tilde{G}_{\sigma}({\rm i}\varepsilon_n) =
\left[
{\rm i}\varepsilon_n+\tilde{\mu}-\epsilon_d - \tilde{\Sigma}_\sigma({\rm i}\varepsilon_n)
- \frac1{\pi}\hskip-3pt \int \hskip-3pt 
d\epsilon^\prime \frac{\Delta_{\rm A}(\epsilon^\prime) }
{{\rm i}\varepsilon_n- \epsilon^\prime}
\right]^{-1},
\end{equation}
where $\tilde{\mu}$ is the chemical potential, $\tilde{\Sigma}_\sigma({\rm i}\varepsilon_n)$ is the self-energy, and
\begin{equation}
\Delta_{\rm A}(\epsilon) =
\frac{\pi}{N_c} \sum_{\bm k} |V_{\bm k}|^2 
\delta\bigl[\epsilon+\tilde{\mu}- E_c({\bm k})\bigr] .
\end{equation}
%
%
The Fermi surface of conduction electrons is defined by $E_c({\bm k})=\tilde{\mu}$. It exists when
\begin{equation}\label{EqFS}
\Delta_{\rm A}(0) > 0 ,
\end{equation}
%
is satisfied. This condition is called the Fermi-surface condition in this paper.

The \mbox{$s$-$d$} model is another effective Hamiltonian for studying the Kondo effect.
According to Yosida's perturbation theory \cite{yosida}
and Wilson's renormalization-group theory, \cite{wilsonKG} 
the ground state is a singlet or a conventional Fermi liquid except when $J_{s\mbox{-}d}=0$, where $J_{s\mbox{-}d}$ is the $s$-$d$ exchange interaction.  Since the \mbox{$s$-$d$} model is derived from the Anderson model, the result for the $s$-$d$ model implies that
the ground state of the Anderson model is also a conventional Fermi liquid.
The Bethe-ansatz solution for the Anderson model confirms that the ground state is a conventional Fermi liquid, at least when $\Delta_{\rm A}(\epsilon)$ is constant. \cite{exact1,exact2,exact3,exact4} 
In general, the nature of the ground state depends only on relevant low-energy properties, such as $\Delta_{\rm A}(0)$, and high-energy properties only  quantitatively renormalize the ground state, as demonstrated by renormalization-group theories for the $s$-$d$ model.\cite{wilsonKG,poorman} 
When the Fermi-surface condition (\ref{EqFS}) is satisfied, therefore, the ground state of the Anderson model is a conventional Fermi liquid except for the case corresponding to $J_{s\mbox{-}d}=0$, i.e., except when $\tilde{U}_\infty/\Delta_{\rm A}(0)=+\infty$ and $n_d=\left<n_{d\uparrow}+n_{d\downarrow}\right>=1$. \\~~\\

\mysection{Formulation} 
\label{SecFormulation}
%
\mysubsection{Fermi liquid as an unperturbed state}
\label{SecUnperturbed}
%
%
Since the superexchange interaction $J$ is antiferromagnetic, only singlet superconductivity is studied in this paper.
When $U_\infty$ is nonzero but finite, $J\ne 0$, or the electron-phonon interaction is present, the single-particle Green function for electrons is given by
\begin{widetext} 
\begin{align}\label{EqGreenNambu0}
{\cal G}_{\sigma}({\rm i}\varepsilon_n,{\bm k}) &= 
%
\left(\begin{array}{cc}
{\rm i}\varepsilon_n -E({\bm k}) + \mu - \Sigma_\sigma({\rm i}\varepsilon_n, {\bm k}) &
-\Delta_{\sigma} ({\rm i}\varepsilon_n,{\bm k}) \\
-\Delta_{\sigma}^* ({\rm i}\varepsilon_n,{\bm k}) & 
{\rm i}\varepsilon_n + E(-{\bm k}) - \mu 
+ \Sigma_{-\sigma}(-{\rm i}\varepsilon_n, -{\bm k})
\end{array} \right) ^{-1} ,
\end{align}
\end{widetext}
%
in the Nambu representation, where $\varepsilon_n=2\pi k_{\rm B}T\left(n+\frac1{2}\right)$ is the fermionic energy, $\Sigma_\sigma({\rm i}\varepsilon_n, {\bm k})$ is the self-energy of electrons, which includes $\Gamma({\rm i}\varepsilon_n)$ defined by eq.~(\ref{EqSigmaL}), and
$\Delta_{\sigma} ({\rm i}\varepsilon_n,{\bm k})$ is the superconducting order parameter; the Green function for phonons is given by
\begin{align}
G_\lambda({\rm i}\omega_\ell,{\bm q})&=
\frac1{[({\rm i}\omega_\ell)^2-\omega_{\lambda{\bm q}}^2]/
(2\omega_{\lambda{\bm q}})- \Sigma_\lambda({\rm i}\omega_\ell,{\bm q})} ,
\end{align}
%
where $\omega_\ell=2\pi\ell k_{\rm B}T$ is the bosonic energy and $\Sigma_\lambda({\rm i}\omega_\ell,{\bm q})$ is the self-energy of phonons.
Because of the inversion symmetry, $E({\bm k})=E(-{\bm k})$, $\Sigma_\sigma({\rm i}\varepsilon_n, {\bm k})=\Sigma_\sigma({\rm i}\varepsilon_n, -{\bm k})$, $\Delta_{\sigma} ({\rm i}\varepsilon_n,{\bm k})=\Delta_{\sigma} ({\rm i}\varepsilon_n,-{\bm k})$, $\omega_{\lambda{\bm q}}=\omega_{\lambda-{\bm q}}$, and $\Sigma_\lambda({\rm i}\omega_\ell,{\bm q})=\Sigma_{\lambda}({\rm i}\omega_\ell,-{\bm q})$.  
The determinant of
${\cal G}_{\sigma}^{-1}({\rm i}\varepsilon_n,{\bm k})$
is given by
\begin{align}\label{EqDet}
\Xi_\sigma({\rm i}\varepsilon_n,{\bm k}) &=
\bigl[ {\rm i}\varepsilon_n -E({\bm k}) + \mu
- \Sigma_\sigma({\rm i}\varepsilon_n, {\bm k}) \bigr]
\nonumber \\ &  \times
\bigl[{\rm i}\varepsilon_n + E({\bm k}) - \mu 
+ \Sigma_{-\sigma}(-{\rm i}\varepsilon_n, {\bm k}) \bigr] 
\nonumber \\ & \quad
- \bigl|\Delta_{\sigma} ({\rm i}\varepsilon_n,{\bm k})\bigr|^2.
\end{align}
Then, the Green function (\ref{EqGreenNambu0}) is also described in such a way that
\begin{equation}\label{EqGreen10}
{\cal G}_{\sigma}({\rm i}\varepsilon_n,{\bm k})=
\left(\begin{array}{cc}
G_{\sigma}({\rm i}\varepsilon_n, {\bm k}) & 
F_{\sigma}({\rm i}\varepsilon_n,{\bm k}) \\
F_{\sigma}^*({\rm i}\varepsilon_n,{\bm k})& 
-G_{-\sigma}(-{\rm i}\varepsilon_n, {\bm k})
\end{array} \right) ,
\end{equation}
where
\begin{equation}\label{EqGreen11}
G_{\sigma}({\rm i}\varepsilon_n, {\bm k})  =
\frac{{\rm i}\varepsilon_n + E(-{\bm k}) - \mu 
+ \Sigma_{\sigma}(-{\rm i}\varepsilon_n, -{\bm k})}
{\Xi_\sigma({\rm i}\varepsilon_n,{\bm k})},
\end{equation}
and
\begin{equation}\label{EqGreen12}
F_{\sigma}({\rm i}\varepsilon_n,{\bm k})  =
\frac{\Delta_{\sigma} ({\rm i}\varepsilon_n,{\bm k})}{
\Xi_\sigma({\rm i}\varepsilon_n,{\bm k})}.
\end{equation}

Feynman diagram in the site representation are classified into single-site and multisite ones.
Vertex corrections due to the random hybridization ${\cal V}$ can be ignored because they are $O(\lambda_0^4)$, with $\lambda_0=\pm 0^{+}$.
The Green function for electrons in the site representation is given by
\begin{align}
G_{ij\sigma}({\rm i}\varepsilon_n) =
\frac1{N}\sum_{\bm k}
e^{{\rm i}{\bm k}\cdot({\bm R}_i-{\bm R}_j)}
G_{\sigma}({\rm i}\varepsilon_n, {\bm k}) ,
\end{align}
for the diagonal component and 
\begin{align}
F_{ij\sigma}({\rm i}\varepsilon_n) =
\frac1{N}\sum_{\bm k}
e^{{\rm i}{\bm k}\cdot({\bm R}_i-{\bm R}_j)}
F_{\sigma}({\rm i}\varepsilon_n, {\bm k}) ,
\end{align}
for the off-diagonal component.
If only the lines of the on-site $U_\infty$ and the site-diagonal $G_{ii\sigma}({\rm i}\varepsilon_n)$ appear in a diagram, the diagram is a single-site one.
If a line of the intersite $J$, the site-off-diagonal $G_{ij\sigma}({\rm i}\varepsilon_n)$ with $i\ne j$, the phonon Green function, or the off-diagonal $F_{ij\sigma}({\rm i}\varepsilon_n)$ appears in a diagram, the diagram is a multisite one.
%
According to this classification, the self-energy of electrons is decomposed into the single-site $\Sigma_\sigma({\rm i}\varepsilon_n)
$ and the multisite $\Delta\Sigma_\sigma({\rm i}\varepsilon_n, {\bm k})$ in such a way that
\begin{equation}
\Sigma_\sigma({\rm i}\varepsilon_n, {\bm k}) =
\Sigma_\sigma({\rm i}\varepsilon_n)
+\Delta\Sigma_\sigma({\rm i}\varepsilon_n, {\bm k})
+ \Gamma({\rm i}\varepsilon_n).
\end{equation}
%

In the Anderson model, all the Feynman diagrams are simply single-site diagrams. 
It is assumed that $\epsilon_d-\tilde{\mu}$ and $\tilde{U}_\infty$ of the Anderson model are exactly the same as $\epsilon_a-\mu$ and $U_\infty$ of the $t$-$J$-$U_\infty$ model, respectively:
%
$\epsilon_d-\tilde{\mu}=\epsilon_d-\tilde{\mu}$ and 
$\tilde{U}_\infty=U_\infty$.
%
The same $U_\infty$ appears in the Feynman diagrams of the Hubbard and Anderson models.
If $\Delta_{\rm A}(\varepsilon)$ is determined to satisfy 
\begin{equation}\label{EqMap-G}
G_{ii\sigma}({\rm i}\varepsilon_n) = \tilde{G}_{\sigma}({\rm i}\varepsilon_n),
\end{equation}
the single-site self-energy is equal to the self-energy for the Anderson model:
$\Sigma_\sigma({\rm i}\varepsilon_n)=\tilde{\Sigma}_\sigma({\rm i}\varepsilon_n)$.
%
It follows from eq.~(\ref{EqMap-G}) that
\begin{equation}\label{EqMap-G1}
\Delta_{\rm A}(\epsilon) =
\text{Im}\left[
\tilde{\Sigma}_\sigma(\epsilon+{\rm i}0)
+ G_{ii\sigma}^{-1}(\epsilon+{\rm i}0)
\right].
\end{equation}
%
Equation~(\ref{EqMap-G}) or (\ref{EqMap-G1}) is the mapping condition to the Anderson model.
In general, $\Delta_{\rm A}(\epsilon)$ depends on $T$, i.e., the mapped Anderson model itself depends on $T$.

According to eq.~(\ref{EqMap-G}), the electron density and the density of states of the $t$-$J$-$U_\infty$ model are the same as those of the Anderson model: $n =n_d$ and
\begin{equation}\label{EqRhoSSA}
\rho (\epsilon) 
= -\frac1{\pi}\mbox{Im} G_{ii\sigma}(\epsilon+{\rm i}0) 
= -\frac1{\pi}\mbox{Im} \tilde{G}_\sigma(\epsilon+{\rm i}0) .
\end{equation}

The mapping condition (\ref{EqMap-G1}) is iteratively treated to obtain the eventual self-consistent $\Delta_{\rm A}(\epsilon)$; not only $\Delta_{\rm A}(\epsilon)$ but also $\tilde{\Sigma}_\sigma(\epsilon)$, $\Delta\Sigma_\sigma(\epsilon + {\rm i}0,{\bm k})$,  $\Delta_\sigma({\rm i}\varepsilon_n,{\bm k})$, and $\Sigma_\lambda(\omega + {\rm i}0,{\bm k})$ should be self-consistently calculated to satisfy eq.~(\ref{EqMap-G1}). 
It is proved in Appendix~\ref{SecProof} that 
\begin{equation}\label{EqSelfDelta}
\Delta_{\rm A}(\epsilon) \ge
-\text{Im} \Gamma (\epsilon + {\rm i}0) .
\end{equation}
According to eqs.~(\ref{EqGamma}) and (\ref{EqSelfDelta}), $\Delta_{\rm A}(0)>0$, i.e.,
the Fermi surface condition (\ref{EqFS}) is satisfied at each step of the iterative process.  Unless $U_\infty/W=+\infty$ and $n=1$ exactly, therefore, the ground state of the Anderson model is a conventional Fermi liquid, so that the single-site self-energy for the ground state is that of the conventional Fermi liquid even if the order parameter is nonzero or the multisite self-energy is anomalous.

We consider the Anderson model in the presence of
an infinitesimally small chemical potential shift $\Delta \mu_{\rm A}$ and an infinitesimally small Zeeman energy $h$. The self-energy $\tilde{\Sigma}_\sigma(\epsilon+{\rm i}0)$ for the Anderson model is expanded in such a way that 
\begin{align}\label{EqExpandSSA}
\tilde{\Sigma}_\sigma(\epsilon + {\rm i}0) &=
\tilde{\Sigma}_0 + \bigl(1-\tilde{\phi}_{\rm e} \bigr)\epsilon
\nonumber \\ & \quad
+ \bigl(1-\tilde{\phi}_c \bigr) \Delta\mu_{\rm A}
+ \bigl(1-\tilde{\phi}_s \bigr)\sigma \tilde{h}
\nonumber \\ & \quad
+ O\bigl(\epsilon^2/k_{\rm B}T_{\rm K}\bigr)
+ O\bigl(k_{\rm B}T^2/T_{\rm K}\bigr), 
\end{align}
where $\tilde{\Sigma}_0$, $\tilde{\phi}_{\rm e}>0$, $\tilde{\phi}_{\rm c}>0$, $\tilde{\phi}_{\rm s}>0$, and $T_{\rm K}$ are all real. Here, $T_{\rm K}$ is the Kondo temperature and $k_{\rm B}T_{\rm K}$ is the energy scale of local quantum spin fluctuations for not only the Anderson model but also the $t$-$J$-$U_\infty$ model.
In the following part of this paper, $T\ll T_{\rm K}$ or
\begin{align}\label{EqTc<<TK}
T_c\ll T_{\rm K},
\end{align}
is assumed, where $T_c$ is the superconducting critical temperature and
$\tilde{\phi}_s  /\tilde{\phi}_{\rm e}$ is denoted by $\tilde{W}_s$:
\begin{equation}
\tilde{W}_s = \tilde{\phi}_s  /\tilde{\phi}_{\rm e},
\end{equation}
which is the so-called Wilson ratio.
In general, $2\tilde{\phi}_{\rm e}=\tilde{\phi}_s+\tilde{\phi}_c$.
When $n\simeq 1$ and $U_\infty/W\gg1$, local spin fluctuations are developed but local charge fluctuations are suppressed, so that $\tilde{\phi}_s \gg 1$ and $\tilde{\phi}_c \ll 1$. Then,
$\tilde{\phi}_{\rm e}\gg 1$
and 
$\tilde{W}_s \simeq 2$.

When the order parameter $\Delta_{\sigma} ({\rm i}\varepsilon_n,{\bm k})$ is ignored, the diagonal part of the Green function is given by
\begin{align}\label{EqGreen20}
G_{\sigma}({\rm i}\varepsilon_n, {\bm k}) &=
\frac1{\tilde{\phi}_{\rm e}}\frac1{
{\rm i}\varepsilon_n -\xi_0({\bm k})-\Delta\bar{\Sigma}_\sigma({\rm i}\varepsilon_n, {\bm k})} 
\nonumber \\ & \quad
+ \mbox{[incoherent term]} ,
\end{align}
where
\begin{equation}\label{EqQuasiDisp0}
\xi_0({\bm k}) = \frac1{\tilde{\phi}_{\rm e}}  \bigl[
\tilde{\Sigma}_0 + E({\bm k} )- \mu \bigr] ,
\end{equation}
and 
\begin{equation}
\Delta\bar{\Sigma}_\sigma({\rm i}\varepsilon_n, {\bm k}) =
\frac1{\tilde{\phi}_{\rm e}} \Delta\Sigma_\sigma({\rm i}\varepsilon_n, {\bm k}) .
\end{equation}
%
The first term in eq.~(\ref{EqGreen20})
is the coherent term, which describes the quasi-particle band with bandwidth $W^*= O(k_{\rm B}T_{\rm K})$ at the top of the lower Hubbard band.
The second term in eq.~(\ref{EqGreen20}) is the incoherent term, which
describes the lower Hubbard band; the upper Hubbard band lies at an infinitely high energy in the limit of $U_\infty/W \to \infty$.

It is convenient to define other renormalized quantities as
\begin{equation}
\bar{\Delta}_{\sigma} ({\rm i}\varepsilon_n,{\bm k}) =
\frac1{\tilde{\phi}_{\rm e}}  \Delta_{\sigma}({\rm i}\varepsilon_n,{\bm k}) ,
\end{equation}
\begin{align}\label{EqDetBar}
\bar{\Xi}_{\sigma}({\rm i}\varepsilon_n,{\bm k}) &=
\bigl[ {\rm i}\varepsilon_n -\xi_0({\bm k})
- \bar{\Delta}\Sigma_\sigma({\rm i}\varepsilon_n, {\bm k})\bigr]
\nonumber \\ & \times
\left[ {\rm i}\varepsilon_n + \xi_0({\bm k})
+ \bar{\Delta}\Sigma_{-\sigma}(-{\rm i}\varepsilon_n,{\bm k})\right]
\nonumber \\ & \quad
- \left|\bar{\Delta}_{\sigma}({\rm i}\varepsilon_n,{\bm k})\right|^2 ,
\end{align}
\begin{equation}\label{EqGreenBar11}
\bar{G}_{\sigma}({\rm i}\varepsilon_n, {\bm k}) =
\frac{{\rm i}\varepsilon_n + \xi_0({\bm k})
+ \Delta \bar{\Sigma}_\sigma (-{\rm i}\varepsilon_n,{\bm k})}
{\bar{\Xi}_{\sigma}({\rm i}\varepsilon_n,{\bm k})}, 
\end{equation}
and
\begin{equation}\label{EqGreenBar12}
\bar{F}_{\sigma}({\rm i}\varepsilon_n,{\bm k}) =
\frac{\bar{\Delta}_{\sigma} ({\rm i}\varepsilon_n,{\bm k})}
{\bar{\Xi}_{\sigma}({\rm i}\varepsilon_n,{\bm k})} .
\end{equation}
When the incoherent part is ignored, the single-particle Green function is given by 
\begin{align}
{\cal G}_{\sigma}({\rm i}\varepsilon_n,{\bm k})=
\frac1{\tilde{\phi}_{\rm e}}
\bar{\cal G}_{\sigma} ({\rm i}\varepsilon_n,{\bm k}), 
\end{align}
where
\begin{align}\label{EqGreenBar10}
\bar{\cal G}_{\sigma} ({\rm i}\varepsilon_n,{\bm k}) &=
\left(\hspace{-3pt}
\begin{array}{cc}
\bar{G}_{\sigma}({\rm i}\varepsilon_n, {\bm k}) & 
\bar{F}_{\sigma}({\rm i}\varepsilon_n,{\bm k}) \\
\bar{F}_{\sigma}^*({\rm i}\varepsilon_n,{\bm k})& 
-\bar{G}_{-\sigma}(-{\rm i}\varepsilon_n, {\bm k})
\end{array} \hspace{-3pt}\right) .
\end{align}

\mysubsection{Intersite exchange interactions}
\label{SecMutualInt}
The polarization function in spin channels is also decomposed into
the single-site $\tilde{\pi}_s({\rm i}\omega_\ell) $ and multisite $\Delta\pi_s({\rm i}\omega_\ell, {\bm q})$:
\begin{equation}\label{EqPiSM}
\pi_s({\rm i}\omega_\ell, {\bm q}) = \tilde{\pi}_s({\rm i}\omega_\ell) 
+\Delta\pi_s({\rm i}\omega_\ell, {\bm q}).
\end{equation}
The spin susceptibilities of the Anderson and $t$-$J$-$U_\infty$
models are given, respectively, by
\begin{equation}\label{EqSusAnd}
\tilde{\chi}_s(i \omega_\ell) = \frac{ 2\tilde{\pi}_s({\rm i}\omega_\ell) }{
 1-U_\infty \tilde{\pi}_s({\rm i}\omega_\ell) } ,
\end{equation} 
and
\begin{equation}\label{EqSusT-J}
\chi_s({\rm i}\omega_\ell, {\bm q})=\frac{2\pi_s({\rm i}\omega_\ell, {\bm q})}{
1-\left[\frac1{4}J_s({\bm q})+  U_\infty \right]\pi_s({\rm i}\omega_\ell, {\bm q})},
\end{equation}
where
\begin{equation}
J_s({\bm q}) = 
\frac1{N}\sum_{\left< ij \right>} J \exp\left[i{\bm q}\cdot
\left({\bm R}_i-{\bm R}_j\right)\right] .
\end{equation}
It should be noted that in the limit of ${U_\infty/W\rightarrow +\infty}$,
\begin{align}\label{EqPiLimit}
U_\infty \tilde{\pi}_s({\rm i}\omega_\ell) \rightarrow 1, 
\end{align}
and
$U_\infty\Delta\pi_s({\rm i}\omega_\ell, {\bm q})
\rightarrow O\left[1/U_\infty \chi_s({\rm i}\omega_\ell, {\bm q})\right]$.
%
A physical picture of Kondo lattices is that local spin fluctuations on different unit cells interact with each other by an intersite exchange interaction. Then, the exchange interaction, which is denoted by $I_s ({\rm i}\omega_\ell, {\bm q})$, is defined by
\begin{equation}\label{EqDefIs}
\chi_s({\rm i}\omega_\ell, {\bm q})=\frac{\tilde{\chi}_s(i \omega_\ell)}
{1- \frac1{4}I_s ({\rm i}\omega_\ell, {\bm q})\tilde{\chi}_s(i \omega_\ell)}.
\end{equation}
It follows from this definition that
\begin{equation}\label{EqIs}
I_s ({\rm i}\omega_\ell, {\bm q}) = J_s({\bm q})+
2 U_\infty^2\Delta\pi_s({\rm i}\omega_\ell, {\bm q}) ,
\end{equation}
in the limit of $U_\infty/W \rightarrow +\infty$.

\begin{figure}
\centerline{
\includegraphics[width=8.0cm]{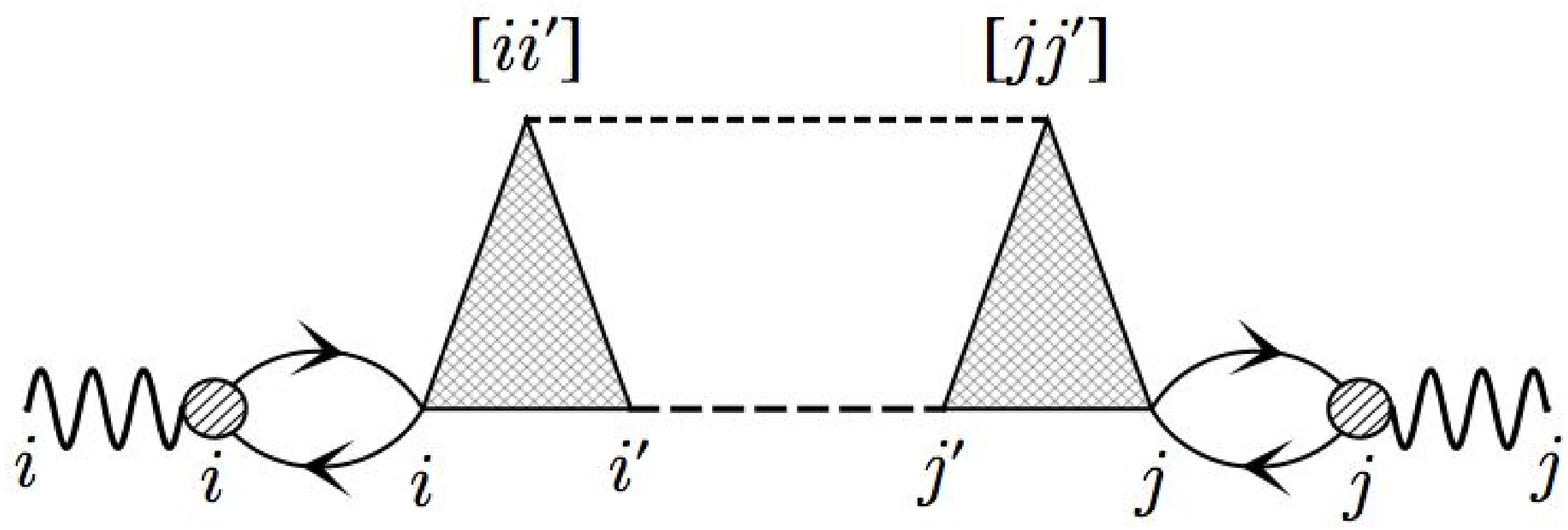}
}
\caption[1]{Feynman diagram  corresponding to $J_\text{sf-ph}({\rm i}\omega_\ell,{\bm q})$.
A wavy solid line denotes $U_\infty$. A hatched circle denotes the irreducible single-site vertex function in spin channels, and
an arrowed solid line denotes the Green function for electrons;
a unit composed of a hatched circle and two arrowed solid lines is the irreducible polarization function in spin channels.
A hatched triangle denotes the electron-phonon interaction,
a dashed line denotes the spin susceptibility, and a dotted line denotes the Green function for phonons.
The $i^\prime$th and $j^\prime$th sites are nearest-neighbor Cu sites 
of the $i$th and $j$th Cu sites, respectively.
The $[ii^\prime]$th and $[jj^\prime]$th sites are O sites between 
the $i$th and $i^\prime$th Cu sites and between the $j$th and $j^\prime$th Cu sites, respectively.
}
\label{fig_el-ph}
\end{figure}
%

The term $2 U_\infty^2\Delta\pi_s({\rm i}\omega_\ell, {\bm q}) $ in eq.~(\ref{EqIs}) is composed of several terms. One is an exchange interaction arising from the virtual exchange of a pair excitation of quasi-particles. When the reducible and irreducible three-point single-site vertex functions in spin channels are denoted by $\tilde{\Lambda}_s({\rm i}\varepsilon_n+{\rm i}\omega_\ell,{\rm i}\varepsilon_n; {\rm i}\omega_\ell)$ and $\tilde{\lambda}_s({\rm i}\varepsilon_n+{\rm i}\omega_\ell,{\rm i}\varepsilon_n; {\rm i}\omega_\ell)$, respectively, it follows that
\begin{equation}
\tilde{\Lambda}_s({\rm i}\varepsilon_n+{\rm i}\omega_\ell,{\rm i}\varepsilon_n; {\rm i}\omega_\ell) =
\frac{
\tilde{\lambda}_s({\rm i}\varepsilon_n+{\rm i}\omega_\ell,{\rm i}\varepsilon_n; {\rm i}\omega_\ell) }
{1 -U_\infty\tilde{\pi}_s({\rm i}\omega_\ell)} , 
\end{equation}
according to the Ward relation. \cite{ward}  Since
$\tilde{\phi}_s=\tilde{\Lambda}_s(0,0; 0)$,
%
it follows that
\begin{equation}\label{Eq3pointVertex}
U_\infty\tilde{\lambda}_s({\rm i}\varepsilon_n+{\rm i}\omega_\ell,{\rm i}\varepsilon_n; {\rm i}\omega_\ell)
=\frac{2 \tilde{\phi}_s}{\tilde{\chi}_s({\rm i}\omega_\ell)} , 
\end{equation}
in the limit of $\varepsilon_n\rightarrow 0$, $\omega_\ell\rightarrow 0$, and $U_\infty/W\rightarrow +\infty$.
When eq.~(\ref{Eq3pointVertex}) is used as an approximation for small $\varepsilon_n$ and $\omega_\ell$ such as $|\varepsilon_n|\ll k_{\rm B}T_{\rm K}$ and $|\omega_\ell|\ll k_{\rm B}T_{\rm K}$,
the exchange interaction is given by
\begin{equation}\label{EqJQ}
J_Q({\rm i}\omega_\ell,{\bm q}) = \frac{4\tilde{W}_s^2}
{\tilde{\chi}_s^2(0)}
\left[P({\rm i}\omega_\ell,{\bm q}) - \tilde{P}({\rm i}\omega_\ell) \right],
\end{equation}
where
\begin{align}\label{EqPolP}
P({\rm i}\omega_\ell,{\bm q}) &=
-2\frac{k_BT}{N} \sum_{\varepsilon_n{\bm k}}
\bigl[\bar{G}_{\sigma}({\rm i}\varepsilon_n+{\rm i}\omega_\ell,{\bm k}+{\bm q} )
\bar{G}_{\sigma}({\rm i}\varepsilon_n,{\bm k})
\nonumber \\ & \quad 
+ \bar{F}_{\sigma}({\rm i}\varepsilon_n+{\rm i}\omega_\ell,{\bm k}+{\bm q} )
\bar{F}_{\sigma}({\rm i}\varepsilon_n,{\bm k})\bigr] ,
\end{align}
and
\begin{equation}
\tilde{P}({\rm i}\omega_\ell) =
- 2 \frac{k_BT}{N^2} \sum_{\varepsilon_n{\bm k}{\bm p}}
\bar{G}_{\sigma}({\rm i}\varepsilon_n+{\rm i}\omega_\ell,{\bm k})
\bar{G}_{\sigma}({\rm i}\varepsilon_n,{\bm p}),
\end{equation}
which is derived in the random-phase approximation (RPA) for a pair excitation of quasi-particles.
In eq.~(\ref{EqJQ}), the single-site term $\tilde{P}({\rm i}\omega_\ell)$ is subtracted to exclude any double counting.
Another term is an exchange interaction arising from the virtual exchange of a coupled excitation of spin fluctuations and phonons, whose exchange process is schematically shown in Fig.~\ref{fig_el-ph}. When eq.~(\ref{EqPiLimit}) is made use of, the exchange interaction is simply given by
\begin{align}\label{EqJsf-ph}
J_\text{sf-ph}({\rm i}\omega_\ell,{\bm q}) &=
2\frac{k_{\rm B}T}{N}\sum_{\omega_n{\bm p}}
K_\lambda^2({\bm q}) D_\lambda({\rm i}\omega_\ell + {\rm i}\omega_n,
2{\bm q}\pm {\bm G} + {\bm p})
\nonumber \\ & \quad \times 
\chi_s({\rm i}\omega_n, {\bm q}+{\bm p}),
\end{align}
where $K_\lambda({\bm q})$ is given by eq.~(\ref{EqCouplSfPh}).
%
When only these terms are considered, it follows that
\begin{equation}\label{EqIs3Terms}
I_s({\rm i}\omega_\ell,{\bm q}) =
J_s({\bm q}) +J_Q({\rm i}\omega_\ell,{\bm q}) 
+J_\text{sf-ph}({\rm i}\omega_\ell,{\bm q}).
\end{equation}
%

When eq.~(\ref{Eq3pointVertex}) is used as an approximation, 
the spin-fluctuation-mediated exchange interaction is given by
\begin{equation}\label{EqChi-J}
\frac1{4} (U \tilde{\lambda}_s)^2
[\chi_s({\rm i}\omega_\ell,{\bm q}) - \tilde{\chi}_s({\rm i}\omega_\ell)]=
\frac1{4} \tilde{\phi}_s^2 I_s^*({\rm i}\omega_\ell,{\bm q}), 
\end{equation}
where $\tilde{\lambda}_s$ denotes $\tilde{\lambda}_s(0,0; 0)$ and  
\begin{align}\label{EqIsStar1}
I_s^*({\rm i}\omega_\ell,{\bm q}) &= 
\frac{ I_s({\rm i}\omega_\ell,{\bm q})}
{1 - \frac1{4} I_s({\rm i}\omega_\ell,{\bm q}) \tilde{\chi}_s({\rm i}\omega_\ell)} .
\end{align}
In eq.~(\ref{EqChi-J}), the single-site term is subtracted to exclude any double counting, and two $\tilde{\phi}_s$ appear as effective three-point vertex functions. 
According to eqs.~(\ref{EqChi-J}) and (\ref{EqIsStar1}), the spin-fluctuation-mediated interaction is 
simply the exchange interaction $I_s^*({\rm i}\omega_\ell,{\bm q})$; the bare  $I_s({\rm i}\omega_\ell,{\bm q})$ is enhanced to $I_s^*({\rm i}\omega_\ell,{\bm q})$ by spin fluctuations.

\mysubsection{Gap equation}
The enhanced exchange interaction $I_s^*({\rm i}\omega_\ell,{\bm q})$ includes no single-site effect from its definition; thus, it does not include the strong effective on-site repulsion between quasi-particles that arises from single-site correlations due to the infinitely large on-site $U_\infty$.
When the effective on-site repulsion is denoted by
$\Gamma_{\uparrow\hskip-1pt\downarrow}$, the gap equation is given by 
\begin{align}\label{EqGapEq}
\bar{\Delta}_\sigma({\rm i}\varepsilon_n,{\bm k}) &=
-\frac{k_BT}{N}\sum_{\varepsilon_{\ell}{\bm p}}
\left[\Gamma_{\uparrow\hskip-1pt\downarrow}
-\frac{3}{4}\tilde{W}_s^2
I_s^*({\rm i}\varepsilon_{\ell}-{\rm i}\varepsilon_n, {\bm p}-{\bm k})\right]
\nonumber \\ & \quad \times 
\frac{\bar{\Delta}_\sigma({\rm i}\varepsilon_{\ell},{\bm p})}{\bar{\Xi}_{\sigma}({\rm i}\varepsilon_{\ell},{\bm p}) } ,
\end{align}
where $\bar{\Xi}_{ \sigma}({\rm i}\varepsilon_n,{\bm k})$ is defined by eq.~(\ref{EqDetBar}). Here, the energy dependence of $\Gamma_{\uparrow\hskip-1pt\downarrow}$ is ignored because its energy scale is $k_{\rm B}T_{\rm K}$, which is assumed to be much larger than $k_{\rm B}T_c$.

When only the first-order term in $I_s^*({\rm i}\omega_\ell,{\bm q})$ is considered, for example, the multisite self-energy of electrons is given by
\begin{align}\label{EqSelf0}
\Delta\bar{\Sigma}_\sigma({\rm i}\varepsilon_n,{\bm k}) &=
-\frac{ k_{\rm B} T}{N} 
\sum_{\varepsilon_{\ell}{\bm p}}e^{{\rm i}\varepsilon_n 0^+}
(-1)\frac{3}{4} \tilde{W}_s^2 
\nonumber \\ & \quad \times
I_s^*({\rm i}\varepsilon_{\ell}-{\rm i}\varepsilon_{n}, {\bm k}- {\bm p}) \bar{G}_\sigma({\rm i}\varepsilon_{\ell},{\bm p}),
\end{align}
where $\bar{G}_{ \sigma}({\rm i}\varepsilon_n,{\bm k})$ is defined by eq.~(\ref{EqGreenBar11}); $\Gamma_{\uparrow\hskip-1pt\downarrow}$ is not included in eq.~(\ref{EqSelf0}) to exclude any double counting.
If superconducting fluctuations develop, their renormalization, which is of higher order in $I_s^*({\rm i}\omega_\ell,{\bm q})$, should be considered.
The self-energy of phonons $\Sigma_\lambda({\rm i}\omega_\ell,{\bm q})$ can also be perturbatively calculated
in ${\cal H}_\text{el-ph}$, as examined in a previous paper.\cite{FJO-elph2}

The theory in this paper is similar to the conventional theory of strong-coupling superconductivity except that the {\it unperturbed} state, which is a conventional Fermi liquid, should be self-consistently constructed with other quantities such as $\bar{\Delta}_{\sigma} ({\rm i}\varepsilon_n,{\bm k})$, $\Delta \bar{\Sigma}_\sigma({\rm i}\varepsilon_n, {\bm k})$, $\Sigma_\lambda({\rm i}\omega_\ell,{\bm q})$, $I_s^*({\rm i}\omega_{\ell}, {\bm q})$, and so forth, to satisfy the mapping condition~(\ref{EqMap-G}) or (\ref{EqMap-G1}). 

\mysection{Application to the Cuprate Oxide}
\label{SecApplication}
\mysubsection{RVB stabilization mechanism}
%
The exchange interaction is expanded in the Fourier series in such a way that
\begin{align}
I_s^*({\rm i}\omega_\ell,{\bm q}) = I_0^*({\rm i}\omega_\ell)
+ 2 \sum_{j} I_j^*({\rm i}\omega_\ell)\eta_{ns}({\bm q}),
\end{align}
where 
\begin{align}
I_j^*({\rm i}\omega_\ell) = \frac1{N}\sum_{\bm q}
I_s^*({\rm i}\omega_\ell,{\bm q})\cos\left[{\bm q}\cdot({\bm R}_j-{\bm R}_0)\right].
\end{align}
Here, ${\bm R}_0$ and ${\bm R}_j$ are the lattice vectors of the origin and its $j$th nearest neighbor, respectively, and $\eta_{ns}({\bm q})$ is the $s$-wave form factor of the $n$th nearest neighbors: 
$\eta_{0s}({\bm q})=1$, $\eta_{1s}({\bm q})$ is defined by eq.~(\ref{EqForm1S}), 
$\eta_{2s}({\bm q})=\cos[(q_x+q_y)a]+\cos[(q_x-q_y)a]$, and so forth.
%
In general,
\begin{align}
I_j^*({\rm i}\omega_\ell) = \left\{\begin{array}{cc}
J + 
\left[I_1^*(0)-J\right]\alpha_1({\rm i}\omega_\ell), & j=1\\
I_j^*(0) \alpha_j({\rm i}\omega_\ell), & j \ne 1 
\end{array}\right. ,
\end{align}
where $\alpha_j({\rm i}\omega_\ell)$ is an analytical function that satisfies 
$\alpha_j(0)=1$ and $\bigl[\alpha_j({\rm i}\omega_\ell)\bigr]_{|\omega_{\ell}|\rightarrow +\infty}=0$.  
When  the constant term $J$ or the superexchange interaction $J_s({\bm q})$ is only considered in eq.~(\ref{EqSelf0}),
the self-energy does not depend on energy; thus, it is simply denoted by $\Delta \bar{\Sigma}_J ({\bm k})$ here. The diagonal part of the Green function is given by $\bar{G}_\sigma({\rm i}\varepsilon_n,{\bm k}) =
1/\left[{\rm i}\varepsilon_n  -\xi({\bm k})\right]$, where
\begin{equation}\label{EqQuasiDisp}
\xi({\bm k}) = \frac1{\tilde{\phi}_{\rm e}}  \bigl[
\tilde{\Sigma}_0 + E({\bm k})  - \mu  \bigr] +
\Delta \bar{\Sigma}_J  ({\bm k}) ,
\end{equation}
is the dispersion relation of renormalized quasi-particles to be determined.
When $T=0$~K, $\tilde{\Sigma}_0- \mu$ is determined so that it satisfy the Fermi-surface sum rule:
\begin{align}
n = \frac{2}{N}\sum_{\bm k}H\hskip-1pt\left[-\xi({\bm k})/W\right] ,
\end{align}
where $H(x)$ is the Heaviside function defined by 
$H(x\ge 0)=1$ and $H(x< 0)=0$. 
It follows that
\begin{equation}\label{EqDeltaSigma}
\Delta \bar{\Sigma}_J ({\bm k}) =
\frac{3}{4} \tilde{W}_s^2 J \eta_{1s}({\bm k})  
\frac1{N}\sum_{\bm p} H\bigl[-\xi({\bm p})/W\bigr]
 \eta_{1s}({\bm p}) , 
\end{equation}
at $T=0$~K.
Since $J$ is antiferromagnetic, the bandwidth of quasi-particles is broadened by this renormalization; even in the limit of $\tilde{\phi}_{\rm e}\rightarrow +\infty$, the bandwidth of quasi-particles is $O(|J|)$. The {\it unperturbed} Fermi liquid constructed in the S$^3$A is further stabilized.
The stabilization is due to the formation of an itinerant local singlet or a resonating valence bond (RVB) on each pair of nearest neighbors and is similar to that in the mean-field RVB theory,\cite{Plain-vanilla} although the stabilized liquid is simply a conventional Fermi liquid. The stabilized Fermi liquid is a more relevant {\it unperturbed} state than the Fermi liquid constructed in the S$^3$A.

The density of states for electrons at the chemical potential is given by
\begin{align}\label{EqRhoRVB}
\rho(0) = \frac1{N}\sum_{\bm k}
\delta\left[\tilde{\Sigma}_0 + E({\bm k})  +
\tilde{\phi}_{\rm e}\Delta \bar{\Sigma}_J  ({\bm k}) - \mu \right] .
\end{align}
It is different from that for quasi-particles, which is defined by
\begin{align}
\rho^*(0)=\frac1{N}\sum_{\bm k}
\delta[\xi({\bm k})] = \tilde{\phi}_{\rm e}\rho(0).
\end{align}
Note that $\rho(0)=O\bigl[1/(\tilde{\phi}_{\rm e}|J|)\bigr]$ in the limit of $\tilde{\phi}_{\rm e} \rightarrow +\infty$.
The density of states $\rho(0)$ is greatly reduced by the existence of $\tilde{\phi}_{\rm e}\Delta \bar{\Sigma}_J  ({\bm k})$ with $\tilde{\phi}_{\rm e} \gg 1 $, although $\rho^*(0)$ is not reduced unless $J$ is large.
According to the Fermi-liquid relation,\cite{Luttinger1,Luttinger2} the specific-heat coefficient is given by $\gamma_C=(2/3)\pi^2k_{\rm B}^2 \rho^*(0)$.
The quasi-particle bandwidth, which is defined by $W^*=W/\tilde{\phi}_{\rm e},$ is estimated to be $W^*\simeq 1/\rho^*(0)\simeq 0.3\mbox{-}0.4$~eV from the observed specific-heat coefficient,\cite{loram} which is, for example, about 10~mJ/mol$\cdot$K$^2$ for optimal-doped cuprate oxides, in which $T_c$ is the highest as a function of doping. 
Since $W=3\mbox{-}4$~eV according to a band calculation, $W/W^*\simeq 10$, which implies that $\tilde{\phi}_{\rm e} \simeq 10$.
On the other hand, the superexchange interaction is as strong as $J=-(0.10\mbox{-}0.15)$~eV.\cite{SuperJ} It is certain that the RVB stabilization mechanism is crucial in the cuprate oxide with $n\simeq 1$.
It will be interesting to experimentally study the reduction of $\rho(0)$ for electrons, not $\rho^*(\epsilon)$ for quasi-particles, in the cuprate oxide with $n\simeq 1$ to obtain evidence that the RVB stabilization mechanism is actually crucial.

If the Kondo temperature is defined by  $k_{\rm B}T_{\rm K} =1/2\rho^*(0)$, then $k_{\rm B}T_{\rm K} \simeq 0.15\mbox{-}0.2$~eV for optimal-doped cuprate oxides and $k_{\rm B}T_{\rm K}\gtrsim 0.2$~eV for over-doped and under-doped cuprate oxides. The condition (\ref{EqTc<<TK}), $T_c\ll T_{\rm K}$, is satisfied in any cuprate-oxide superconductor.

\mysubsection{Kinks in the normal state}
\label{SecKinks}
The exchange interaction 
$I_s^*({\rm i}\omega_\ell,{\bm q})$ or its Fourier component $I_j^*({\rm i}\omega_\ell)$ should be self-consistently calculated using the self-energies of electrons and phonons, the order parameter, the spin susceptibility, and so forth. Since it is a difficult task to complete the self-consistent process, a phenomenological theory is developed in the following part of this paper; each crucial effect is studied in a non-self-consistent manner or independently by assuming that the analytical continuation of $\alpha_j({\rm i}\omega_\ell)$ onto the real axis can be approximately given by 
\begin{align}\label{EqAlpha}
\alpha_j(\omega \pm {\rm i}0)= \frac{\nu_j^2+\gamma_j^2}
{2\nu_j}\left[
\frac{1}{\omega+\nu_j \pm {\rm i}\gamma_j} - \frac{1}{\omega-\nu_j \pm {\rm i}\gamma_j}
\right],
\end{align}
where $\nu_j$ and $\gamma_j$ are positive constants.  
Figure~\ref{fig_kink}(a) shows $\alpha_j(\omega+{\rm i}0)$ for several $\gamma_j/\nu_j$.
According to the analysis in Appendix~\ref{SecDynamicalSus},
it is reasonable to assume that $\nu_j\simeq \omega_{\rm ph}$ in the normal state and that
$(1/2)\epsilon_{\rm G} \lesssim \nu_j \lesssim (1/2)\epsilon_{\rm G} + \omega_{\rm ph}$ in the superconducting state, where $\omega_{\rm ph}$ is the energy of relevant phonons and $\epsilon_{\rm G}\simeq \max[2|\bar{\Delta}_{\sigma}(+{\rm i}0,{\bm k})|]$.
For the sake of simplicity, it is also assumed that the density of states for quasi-particles in the {\em unperturbed} state is constant, so that 
\begin{align}\label{EqConstRho}
\rho^*(\epsilon)=
\frac1{N}\sum_{\bm k} \delta\bigl[\epsilon-\xi({\bm k})\bigr]
= \rho^*_0 H\hskip-2pt\left[1-|\epsilon|/D^*\right] .
\end{align}
Here, $D^*\simeq W/2\tilde{\phi}_{\rm e}$ is half the quasi-particle bandwidth and $\rho^*_0=1/(2D^*)$.

Since the self-energy due to the on-site component does not depend on ${\bm k}$, it is simply denoted by $\Delta\bar{\Sigma}_{0}({\rm i}\varepsilon_n)$ here; the retardation effect is crucial in the on-site effect. 
When $\bar{G}_\sigma({\rm i}\varepsilon_n,{\bm k})=1/[{\rm i}\varepsilon_n - \xi({\bm k})]$ is used in eq.~(\ref{EqSelf0}), 
$\Delta\bar{\Sigma}_{0}({\rm i}\varepsilon_n)$ is given by
\begin{align}\label{EqSelfNormal0}
\Delta \bar{\Sigma}_{0}({\rm i}\varepsilon_n) &=
g_0 \int_{-D^*}^{D^*} \hskip-5pt d\epsilon\hskip1pt
k_{\rm B} T \sum_{\varepsilon_\ell}
\frac{\alpha_0({\rm i}\varepsilon_\ell-{\rm i}\varepsilon_n)} 
{{\rm i}\varepsilon_\ell-\epsilon},
\end{align}
where
\begin{align}
g_0= \frac{3}{4} \tilde{W}_s^2 \rho^*_0 I_0^*(0) .
\end{align}
Since $(1/N)\sum_{\bm q}J_s({\bm q})=0$ and $(1/N)\sum_{\bm q}J_Q({\rm i}\omega_\ell,{\bm q})=0$,  $I_0^*(0)$ is simply given by
\begin{equation}
I_0^*(0)= \frac1{N}\sum_{\bm q}\left[
J_\text{sf-ph}(0,{\bm q})+\frac1{4}I_s^2(0,{\bm q})
\chi_s(0,{\bm q})\right].
\end{equation}
It is obvious that $I_0^*(0)$ is positive and that $g_0>0$.
It should be noted that ${\rm Re}\Delta \bar{\Sigma}_{0}(\epsilon+{\rm i}0)$  and ${\rm Im}\Delta\bar{\Sigma}_{0}(\epsilon+{\rm i}0)$
are odd and even functions of $\epsilon$, respectively, within the phenomenological model used in this subsection.

In this subsection, $\nu_0/D^*=1/4$ is assumed; results for low-energy parts are not sensitive to the value of $\nu_0/D^*$ provided that $\nu_0/D^*$ is sufficiently small.
Figure~\ref{fig_kink}(b) shows $\Delta \bar{\Sigma}_{0}(\epsilon+{\rm i}0)/(g_0\nu_0)$ for $\gamma_0/\nu_0=0.1$, 0.2, 0.4, and 0.8.
The renormalized dispersion relation of quasi-particles, which is denoted by $\xi^*({\bm k})$, is approximately determined by
\begin{align}
\xi^*({\bm k}) = \xi({\bm k}) + {\rm Re} \left\{\Delta\bar{\Sigma}_{0}\left[\xi^*({\bm k})+{\rm i}0\right]\right\}.
\end{align}
Figure~\ref{fig_kink}(c) shows $\xi^*({\bm k})$ as a function of $\xi({\bm k})$.
Since $\xi({\bm k})$ is a smooth function of ${\bm k}$, $\xi^*({\bm k})$ shows kinks as a function of ${\bm k}$ at approximately $\pm\nu_0$ above and below the chemical potential.
Figure~\ref{fig_kink}(d) shows the density of states for renormalized quasi-particles:
\begin{align}
\rho^*(\epsilon) &= -\frac{\rho^*_0}{\pi} \hskip1pt{\rm Im}
\int_{-D^*}^{+D^*} \hskip-8pt d\epsilon^{\prime}
\frac1{\epsilon - \epsilon^{\prime} - \Delta\bar{\Sigma}_{0}(\epsilon+{\rm i}0)}.
\end{align}
When a kink appears at approximately $\epsilon\simeq \pm \nu_0$ in the dispersion relation $\xi^*({\bm k})$,
a small structure appears at approximately $\epsilon\simeq \pm \nu_0$ in 
the density of states $\rho^*(\epsilon)$.

\begin{figure*}
\centerline{
\begin{minipage}{8cm}
\includegraphics[width=7.cm]{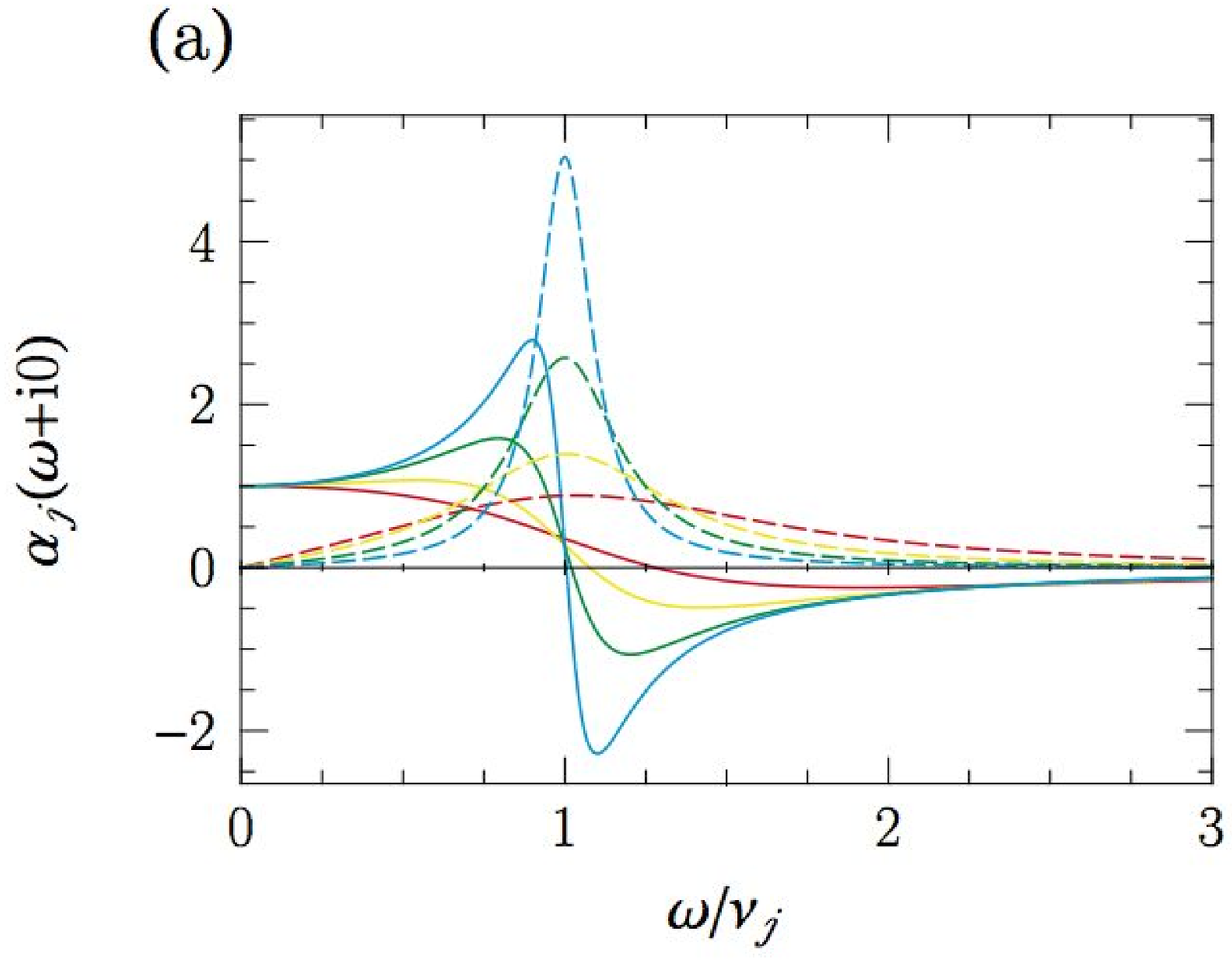}
\end{minipage}
\begin{minipage}{8cm}
\includegraphics[width=7.cm]{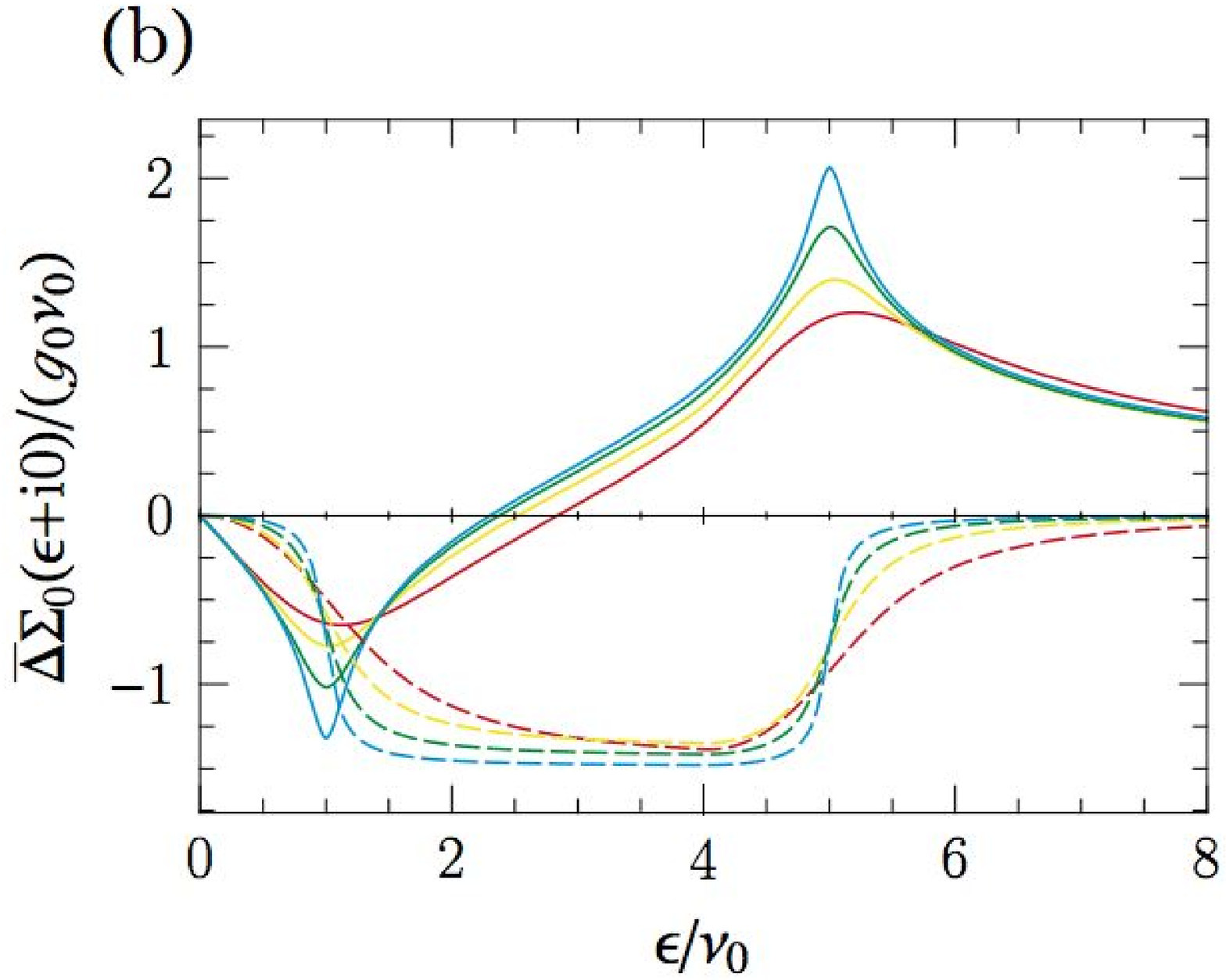}
\end{minipage}
}
\centerline{
\begin{minipage}{8cm}
\includegraphics[width=7.cm]{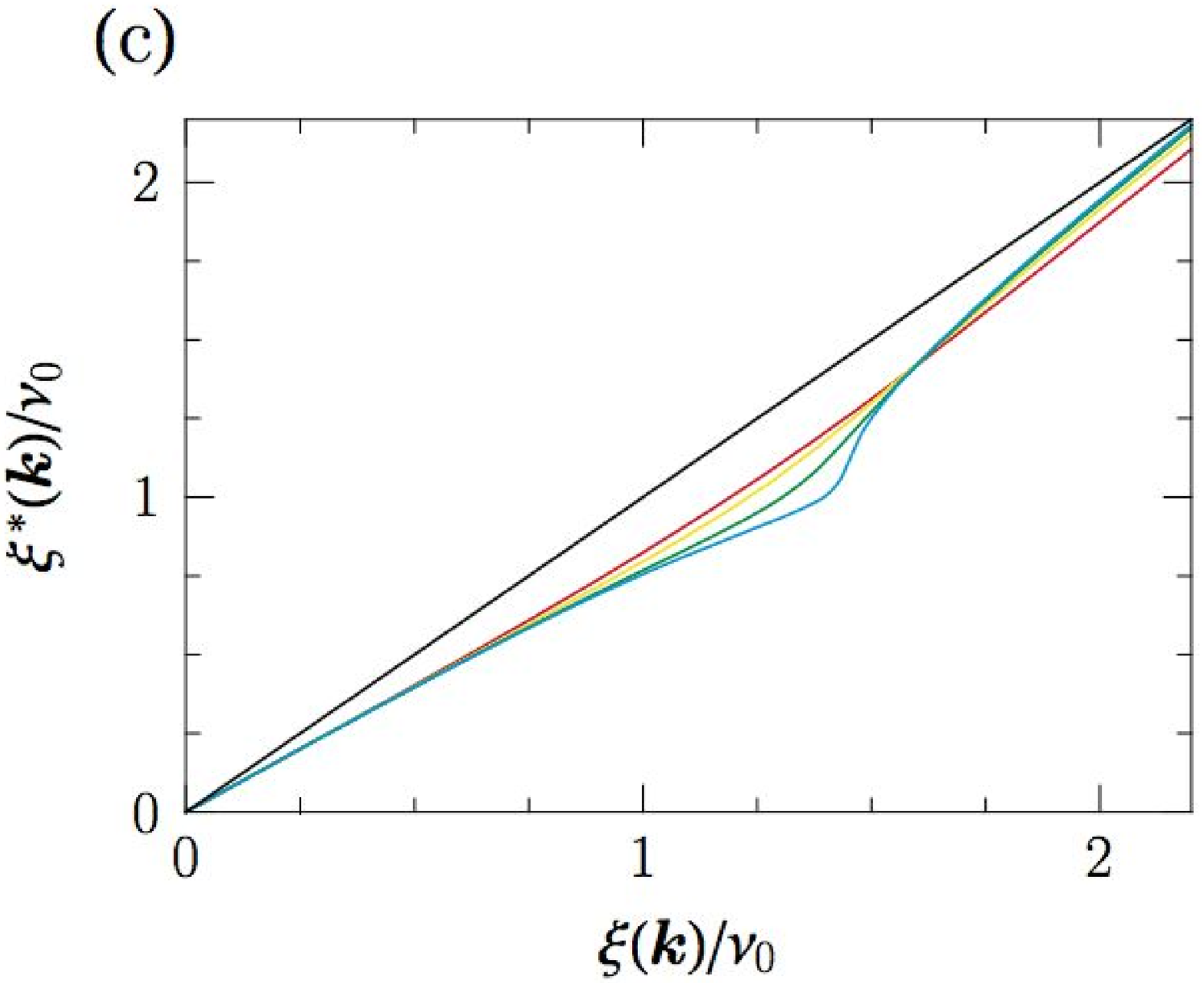}
\end{minipage}
\begin{minipage}{8cm}
\includegraphics[width=6.5cm]{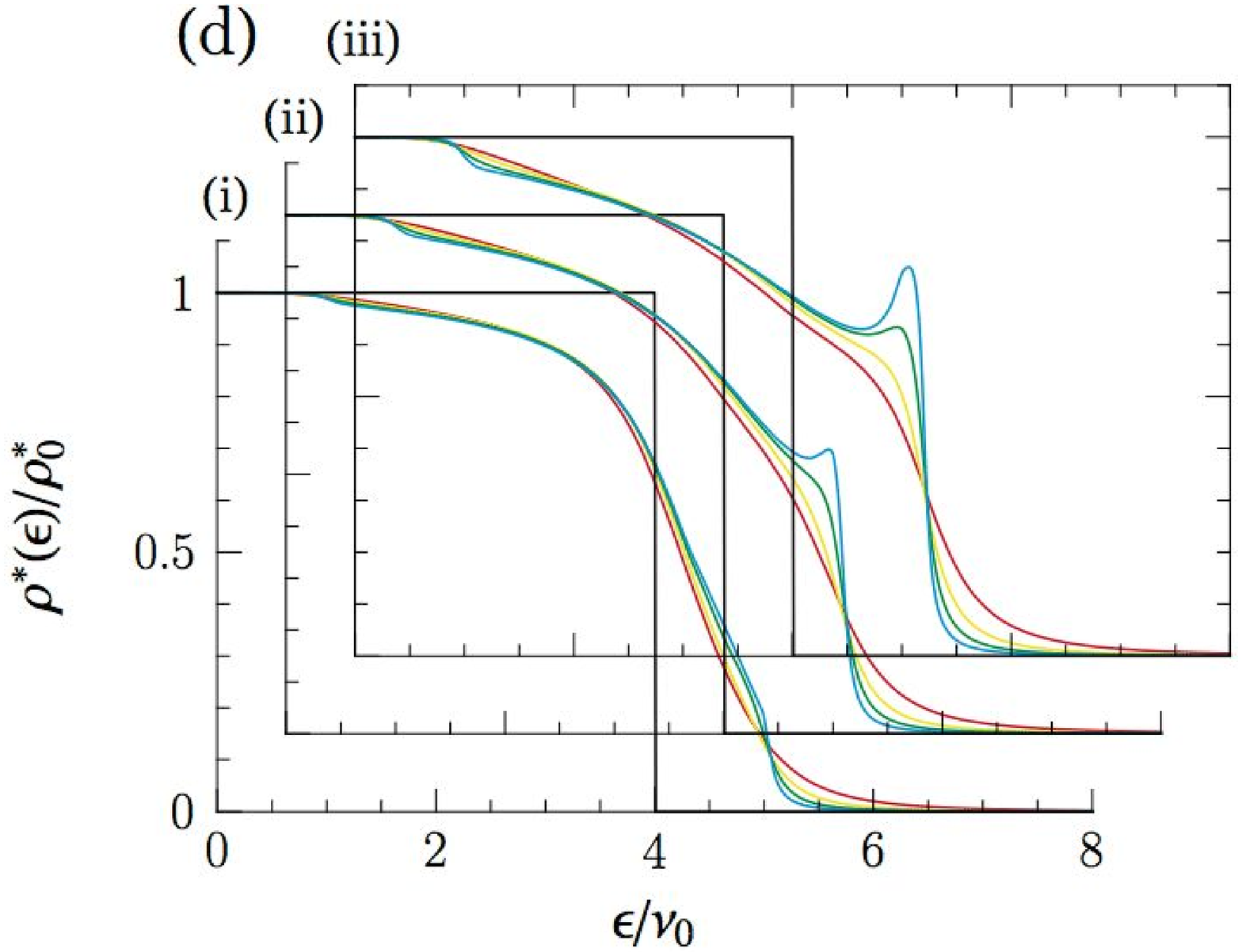}
\end{minipage}
}
\caption[1]{
(a) Phenomenological function $\alpha_j(\omega+{\rm i}0)$,
(b) self-energy 
$\Delta\bar{\Sigma}_{0}(\epsilon+{\rm i}0)$ for $D^*/\nu_0=4$,
(c) kink structure for $g_0=0.3$ and $D^*/\nu_0=4$,
and (d) density of states $\rho^*(\epsilon)$ for $D^*/\nu_0=4$ and three values of $g_0$: (i) $g_0=0.3$, (ii) $0.5$, and (iii) $0.7$. In each figure, (red) $\gamma_j/\nu_j=0.8$, (yellow) $0.4$, (green) $0.2$, and (cyan) $0.1$. In (a) and (b), the solid and dashed lines show the real and imaginary parts, respectively. In (c), the relation $\xi^*({\bm k}) = \xi({\bm k})$ is shown by a black line for comparison. In (d), $\rho^*(\epsilon)$ for $g_0=0$ is shown by a black line for each value of $g_0$ for comparison.
}
\label{fig_kink}
\end{figure*}


\mysubsection{Dip-and-hump structure in the superconducting state}
\label{SecDipHump}
When the on-site $\Gamma_{\uparrow\hskip-1pt\downarrow}$ and the nearest-neighbor $I_1^*({\rm i}\omega_\ell)$ are only considered, only $s$ and $d_{x^2-y^2}$ waves are possible.
Since $\Gamma_{\uparrow\hskip-1pt\downarrow}$ is strongly repulsive and $\eta_{1d}^2({\bm k})\ll 1$ on the Fermi surface for $n\simeq 1$, when $n\simeq 1$, $T_c$ of the $s$ wave is much lower than $T_c$ of the $d_{x^2-y^2}$ wave.
Thus, the $d_{x^2-y^2}$ wave is only considered here.
Since $\Gamma_{\uparrow\hskip-1pt\downarrow}$ has no effect on the $d_{x^2-y^2}$ wave,
the order parameter $\bar{\Delta}_\sigma({\rm i}\varepsilon_n,{\bm k})$ is decomposed into the form factor $\eta_{1d}({\bm k})$ and an energy-dependent part, which is called a gap function and is simply denoted by $\bar{\Delta}_{\rm G}({\rm i}\varepsilon_n)$ here:
\begin{align}\label{EqGapFuncDef}
\bar{\Delta}_\sigma({\rm i}\varepsilon_n,{\bm k})=
\frac1{2}\eta_{1d}({\bm k}) \bar{\Delta}_{\rm G}({\rm i}\varepsilon_n). 
\end{align}
To simplify the numerical processes, the renormalization of quasi-particles studied in \mbox{\S\hskip2pt\ref{SecKinks}} is ignored, and
\begin{align}\label{EqForm1dCos}
\eta_{1d}({\bm k})= 2\cos(2\varphi), 
\end{align}
where
$\varphi=\tan^{-1}(k_y/k_x)$, is used as an approximation instead of
eq.~(\ref{EqForm1D}). 
The particle-hole symmetry does not exist in the original $t$-$J$ model, in general. However, the particle-hole asymmetry is not crucial here, similarly to in the BCS theory. Thus, the constant and symmetric density of states given by eq.~(\ref{EqConstRho}) is assumed.
When the summation along the imaginary axis is transformed into
the integration along the real axis in eq.~(\ref{EqGapEq}), the gap equation is given by
\begin{align}\label{EqGapEqOnRealAxis}
\bar{\Delta}_{\rm G}& (\epsilon+{\rm i}0)  
\nonumber \\ & \quad 
= 
- g_1  \int_{-\infty}^{+\infty} \hskip-12pt d\xi 
\int_{0}^{2\pi} \frac{d\varphi}{2\pi}2\cos^2(2\varphi)
\nonumber \\ & \qquad \times
\int_{-\infty}^{0} \hskip-2pt
\times \frac{dz}{\pi} \Biggl\{{\rm Im}\left[
\frac{I_1^*(z+{\rm i}0)}{I_1^*(0)} \right]
\frac{\bar{\Delta}_{\rm G}(z+\epsilon+{\rm i}0)}
{\bar{\Xi}(z+\epsilon+{\rm i}0,\xi,\varphi)} 
\nonumber \\ & \qquad
+  \frac{I_1^*(z+\epsilon+{\rm i}0)}{I_1^*(0)} 
{\rm Im}\left[ \frac{\bar{\Delta}_{\rm G}(z+{\rm i}0)}
{\bar{\Xi}(z+{\rm i}0,\xi,\varphi)} \right] \Biggr\},
\end{align}
at $T=0$~K, where
\begin{align}
g_1 = - \frac{3}{2} \left(\frac{\tilde{\phi}_s}{\tilde{\phi}_{\rm e}}\right)^2\rho^*_0 I_1^*(0),
\end{align}
is the dimensionless coupling constant and
\begin{align}
\bar{\Xi}(\epsilon+{\rm i}0, \xi,\varphi) = 
(\epsilon+{\rm i}0)^2 -\left[\xi^2 + \cos^2(2\varphi)\left|\bar{\Delta}_{\rm G}(\epsilon+{\rm i}0)\right|^2\right].
\end{align}
In eq.~(\ref{EqGapEqOnRealAxis}), the integration over $\xi$ is extended from $-D^*\le \xi \le +D^*$ to $-\infty< \xi <+\infty$ to simplify the numerical processes. Instead of this, a cutoff is introduced into the superexchange interaction such that the constant $J$ is replaced by $J \alpha_J(\omega+{\rm i}0)$, where
\begin{align}\label{EqCutoffModel1}
\alpha_J(\omega+{\rm i}0) = H(1- |\omega|/\theta_J) .
\end{align}
Here, $\theta_J$ is the phenomenological cutoff parameter; $\nu_1/\theta_J=1/4$ or $\theta_J/\nu_1=4$ is assumed.
Although the analyticity of the summand or integrand is assumed in the transformation from the summation along the imaginary axis to the integration along the real axis, 
this cutoff function is not analytical. 
To confirm the relevance of the cutoff model of eq.~(\ref{EqCutoffModel1}), another cutoff model is also studied:
$\alpha_J(\omega+{\rm i}0)=H_{\gamma_J}(\omega)/H_{\gamma_J}(0)$, where
\begin{align}
H_{\gamma_J}(\omega) &=
\frac1{{\rm i} \pi}\ln\frac{\omega-\theta_J +{\rm i} \gamma_J}
{\omega+ \theta_J + {\rm i}\gamma} ,
\end{align}
is an analytical function that satisfies $\left[H_{\gamma_J}(\omega)\right]_{\gamma_J\rightarrow +0}=H(1- |\omega|/\theta_J)$.
Here, $\gamma_J=0.1\theta_J$ is assumed
to remove the logarithmic singularity in ${\rm Im}H_{\gamma_J}(\omega)$.
These two models are called the non-analytical and analytical models, respectively; the main difference between them is that ${\rm Im}\alpha_J(\omega+i0)=0$ in the non-analytical model but ${\rm Im}\alpha_J(\omega+i0)\ne 0$ in the analytical model.
When $J/W<0$, $I_s^*(0,{\bm q})$ is antiferromagnetic or $I_s^*(0,{\bm q})$ has its maximum near ${\bm Q}_M=\left(\pm1, \pm1\right)(\pi/a)$. Then, $I_1^*(0)/W<0$ so that $g_1>0$.
When a constant $r_1$ is defined by $r_1 = J/I_1^*(0)$, 
$I_1^*(\omega+{\rm i}0)/I_1^*(0)$, which appears in the gap equation (\ref{EqGapEqOnRealAxis}), is given by
\begin{align}
I_1^*(\omega+{\rm i}0)/I_1^*(0)=
r_1\alpha_J(\omega+{\rm i}0) + (1-r_1)\alpha_1(\omega+{\rm i}0).
\end{align}

\begin{table}
\caption{
Adjusted $g_1$ for the analytical (A) and non-analytical (non-A) models with $\gamma_1/\nu_1=0.1$ to reproduce $2\bar{\Delta}_{\rm G}(0)/\nu_1=1$ or  $2\bar{\Delta}_{\rm G}(0)/\theta_J=1/4$.
}
\label{tb-ajustedg0}
\begin{center}
\begin{tabular}{|c||c|c|}
\hline
$r_1$ & ~~~~A~~~~ & ~~non-A~~ \\
\hline\hline
0.3 & 0.5455 & 0.6080 \\
\hline
0.5 & 0.4745 & 0.5175 \\
\hline
0.7 & 0.4323 & 0.4605 \\
\hline
1 & 0.3893 & 0.3928 \\
\hline
\end{tabular}
\end{center}
\end{table}

\begin{figure}
\centerline{
\includegraphics[width=7.0cm]{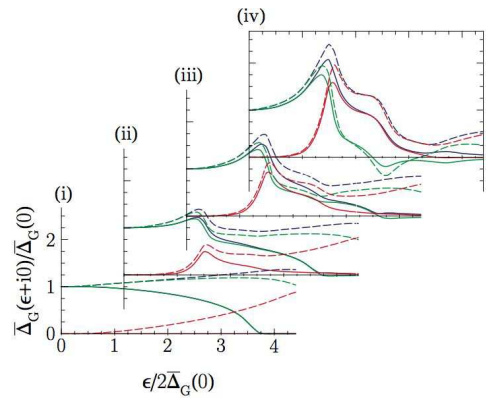}
}
\caption[1]{Gap function $\bar{\Delta}_{\rm G}(\epsilon+{\rm i}0)$ for  $\theta_J/\nu_1=4$ and $\gamma_1/\nu_1=0.1$: (solid line)
the non-analytical model and (dashed line) the analytical model;  
(green) ${\rm Re}\left[\bar{\Delta}_{\rm G}(\epsilon+{\rm i}0)\right]$, 
(red) ${\rm Im}\left[\bar{\Delta}_{\rm G}(\epsilon+{\rm i}0)\right]$, and (blue) $\left|\bar{\Delta}_{\rm G}(\epsilon+{\rm i}0)\right|$;
(i) $r_1=1$, (ii) $r_1=0.7$, (iii) $r_1=0.5$, and (iv) $r_1=0.3$.
In (i), the imaginary part is vanishingly small in the non-analytical model.
}
\label{fig_gap}
\end{figure}

Since the phase of $\bar{\Delta}_{\rm G}(\epsilon+{\rm i}0)$ is arbitrary, it is chosen in such a way that  $\bar{\Delta}_{\rm G}(+{\rm i}0)$ is real and positive; thus,
it is simply denoted by $\bar{\Delta}_{\rm G}(0)$ here.
Since the particle-hole symmetry exists in the simplified phenomenological model, 
the real and imaginary parts of $\bar{\Delta}_{\rm G}(\epsilon+{\rm i}0)$ are even and odd functions, respectively.
Figure~\ref{fig_gap} shows $\bar{\Delta}_{\rm G}(\epsilon+{\rm i}0)$ as a function of $\epsilon$ for the two models with $\gamma_1/\nu_1=0.1$.
To compare the non-analytical and analytical models, the dimensionless coupling constant $g_1$ is adjusted such that 
\begin{align}\label{EqAdjustmnet}
2\bar{\Delta}_{\rm G}(0)/\nu_1=1 ,
\end{align}
or $2\bar{\Delta}_{\rm G}(0)/\theta_J=1/4$.
The adjusted values of $g_1$ for $\gamma_1/\nu_1=0.1$ are shown in Table~\ref{tb-ajustedg0}.
%
%
The gap function $|\bar{\Delta}_{\rm G}(\epsilon+{\rm i}0)|$ is larger in the analytical model than in the non-analytical model because of ${\rm Im}\alpha_J(\omega\pm {\rm i}0)$, but no crucial difference can be seen  between them in the low-energy region of $|\epsilon|< (1.5\mbox{-}2.0) 2\bar{\Delta}_{\rm G}(0)$.
Results for only the non-analytical model are shown in the following part of this paper. The adjustment of eq.~(\ref{EqAdjustmnet}) for $g_1$ is made in each case.

\begin{figure}
\centerline{
\includegraphics[width=7.0cm]{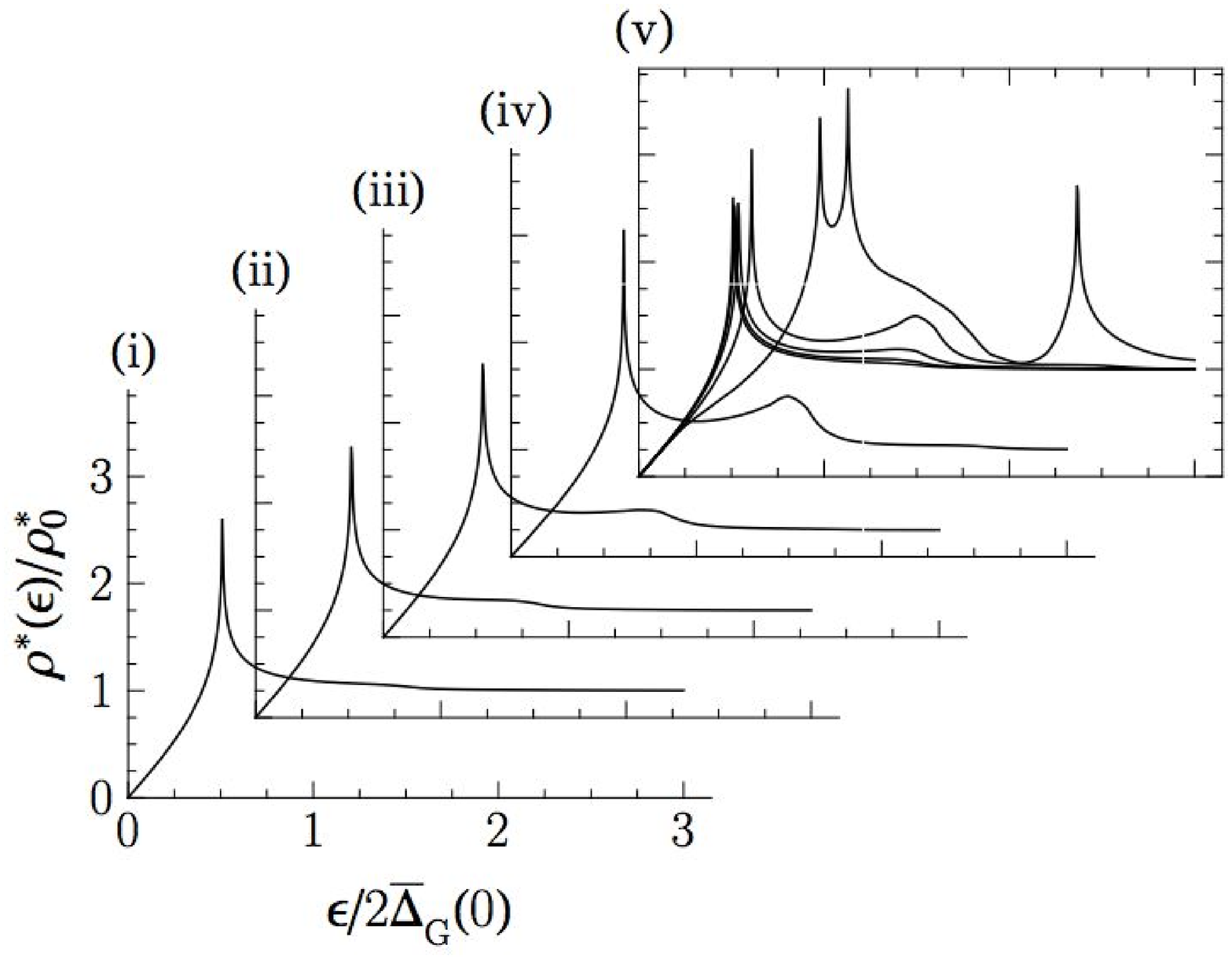}
}
\caption[1]{Density of states $\rho^*(\epsilon)$ for $\theta_J/\nu_1=4$ and $\gamma_1/\nu_1=0.1$: (i) $r_1=0.7$, (ii) $r_1=0.5$, (iii) $r_1=0.3$, (iv) $r_1=0.2$, and (v) $r_1=0.19$; for comparison, $\rho^*(\epsilon)$ for  $r_1=0.7$, 0.5, 0.3, and 0.2 are also shown in (v).
As can be seen in (v),
$\rho^*(\epsilon)$ hardly depends on $r_1$ in the low-energy region of $|\epsilon|/2\bar{\Delta}_{\rm G}(0)\lesssim 0.25$ or $|\epsilon|/\bar{\Delta}_{\rm G}(0)\lesssim 0.5$ because $\bar{\Delta}_{\rm G}(0)/\nu_1=0.5$ is assumed in every case, but 
the apparent size of the gap for $r=0.19$ is about twice as large as that for  $0 \le r_1 \le 0.2$.
In the case of $r_1=0.19$, the sum rule for the density of states $\rho^*(\epsilon)$ is significantly violated 
because of non-analyticity involved in the non-analytical model; thus, the sharp three-peak structure for $r_1=0.19$ should not be seriously considered and should be regarded as an approximate or qualitative one.
}
\label{fig_rho}
\end{figure}
\begin{figure}
\centerline{
\includegraphics[width=7.0cm]{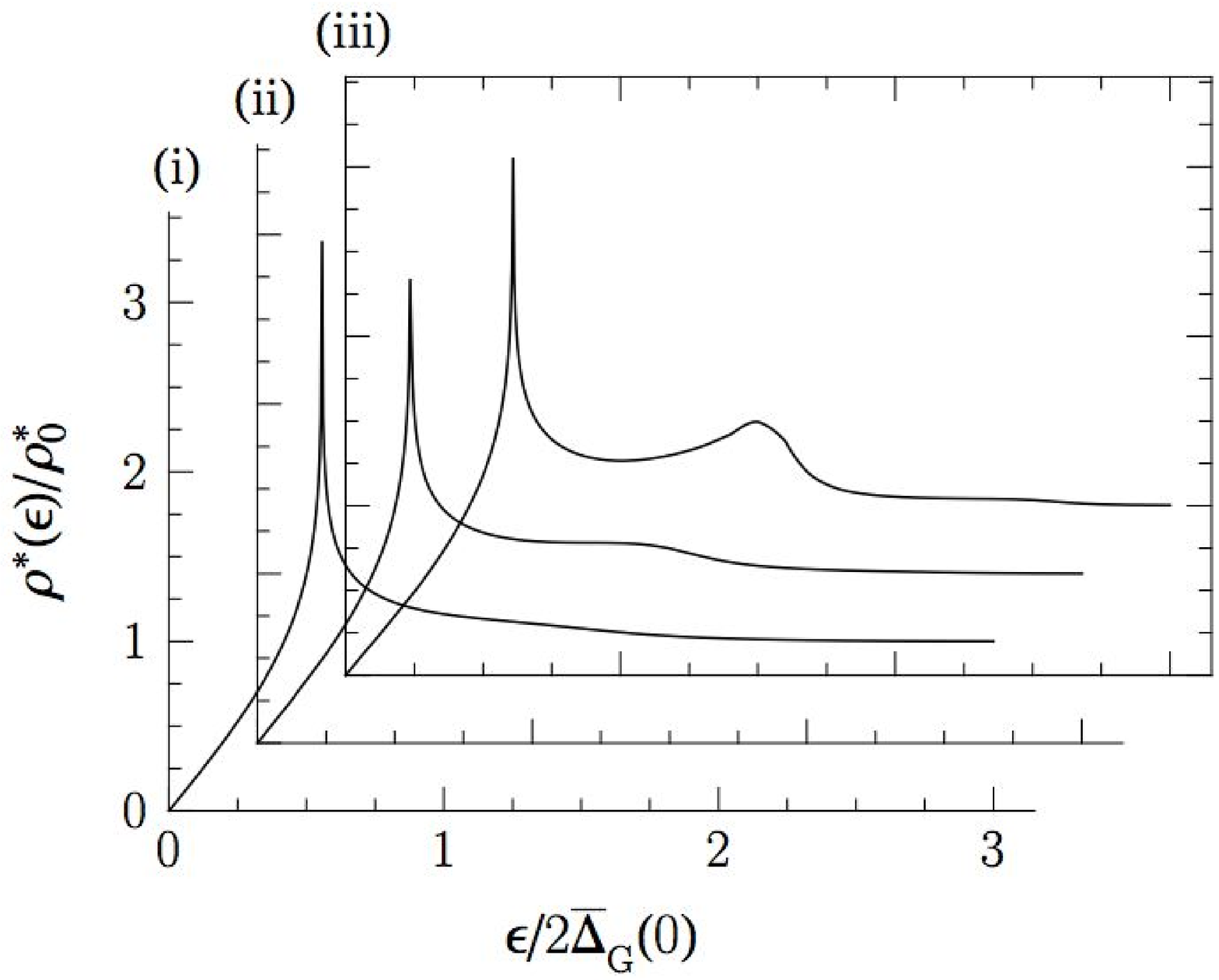}
}
\caption[1]{
Density of states $\rho^*(\epsilon)$ for $\theta_J/\nu_1=4$ and
$r_1=0.2$: (i) $\gamma_1/\nu_1=0.4$, (ii) $0.2$, and (iii) $0.1$.
}
\label{fig_gap-eta}
\end{figure}

%
Figure~\ref{fig_rho} shows the density of states for Bogoliubov's quasi-particles, which is given by
\begin{align}\label{EqRho*}
\rho^*(\varepsilon) = 
\rho_0^*\int_{-\infty}^{+\infty} \hskip-8pt d\xi
\int_{0}^{2\pi} \frac{d\varphi}{2\pi}
\left(-\frac1{\pi}\right) {\rm Im}\frac{\varepsilon + \xi+{\rm i}\gamma_Q}
{\bar{\Xi}(\varepsilon+{\rm i}\gamma_Q, \xi,\varphi) },
\end{align}
where 
$\gamma_Q/\theta_J\rightarrow +0$ is assumed here. 
%
The coherence peak appears at $\epsilon\simeq \bar{\Delta}_{\rm G}(0)$ 
for $r_1\ge 0.2$.
A dip and a hump appear at $\epsilon\simeq \left[\nu_1+\bar{\Delta}_{\rm G}(0)\right]/2$ and $\epsilon\simeq \nu_1$, respectively, for $0.2 \le r_1 \lesssim 0.3$.
For $r_1=0.19$, on the other hand, three peaks appear, which implies that a first-order transition or a sharp crossover occurs between $r_1> r_c$ and $r_1 < r_c$, with $r_c\simeq 0.2$. 
The transition or crossover is studied in the next subsection of \mbox{\S\hskip2pt\ref{SecCrossover}}.
Figure~\ref{fig_gap-eta} shows the density of states for $r_1=0.2$ and three values of $\gamma_1/\nu_1$: 0.1, 0.2, and 0.4.
The dip-and-hump structure can only be seen for sufficiently small $\gamma_1/\nu_1$.

According to the analysis in Appendix~\ref{SecDynamicalSus}, it is likely that $\bar{\Delta}_{\rm G}(0) \lesssim \nu_1\lesssim \bar{\Delta}_{\rm G}(0)+\omega_{\rm ph}$.
The observed dip-and-hump structure\cite{dip-hump} is consistent with results in this paper.

\begin{figure}
\centerline{
\includegraphics[width=7.0cm]{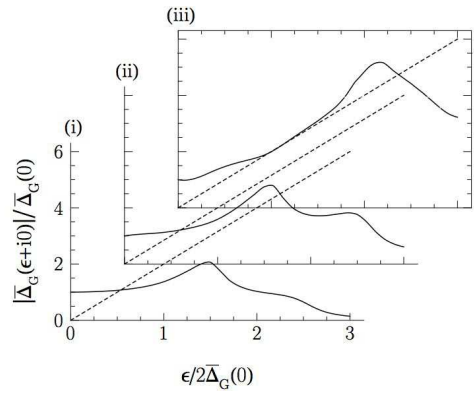}
}
\caption[1]{
Gap function $|\bar{\Delta}_{\rm G}(\epsilon+{\rm i}0)|$ for $\theta_J/\nu_1=4$ and $\gamma_1/\nu_1=1$: (i) $r_1=0.3$, (ii) $0.2$, and (iii) $0.19$.
The relation $|\bar{\Delta}_{\rm G}(\epsilon+{\rm i}0)|=\epsilon$ is shown by a dotted line. 
}
\label{fig_gapCrt}
\end{figure}

\mysubsection{Small-gap and large-gap phases}
\label{SecCrossover}
The dispersion relation of quasi-particles in the superconducting state, which is denoted by $\xi_{\pm}^* ({\bm k})$, is also defined by the pole of the Green function. It is approximately given by a solution of 
\begin{align}\label{EqXi*Sol}
\xi_{\pm}^* ({\bm k}) = 
\pm \sqrt{\xi^2 ({\bm k})+ \frac1{2^2}
\eta_{1d}^2({\bm k}) \left| \bar{\Delta}_{\rm G}
\left[\xi_{\pm}^* ({\bm k}) +{\rm i}0\right] \right|^2}.
\end{align}
According to this equation, an effective gap as a function of $\varphi$, which is denoted by
$\epsilon_{\rm G}(\varphi)$, is also defined by a solution of 
\begin{align}\label{EqEffectGap}
\mbox{$\frac1{2}$}\epsilon_{\rm G}(\varphi) = 
\left| \cos(2\varphi) \bar{\Delta}_{\rm G}\left[\mbox{$\frac1{2}$}
\epsilon_{\rm G}(\varphi)+{\rm i}0\right] \right|,
\end{align}
which is the solution of eq.~(\ref{EqXi*Sol}) for ${\bm k}$ satisfying $\xi ({\bm k})=0$; eqs.~(\ref{EqConstRho}) and (\ref{EqForm1dCos}) are assumed.  As is implied by Fig.~\ref{fig_gapCrt},
there is only a single solution when $r_1\le 0.2$ while
there are three solutions when $r_1=0.19$ and $|\varphi|$ is small.
Because of the appearance of extra poles in the Green function, the gap equation gives a larger gap function when $r_1<r_c\simeq 0.2$ than when $r_1>r_c\simeq 0.2$. 
The jump between $r_1>r_c$ and $r_1<r_c$ is that between a small-gap phase for $r_1>r_c$ and a large-gap phase for $r_1<r_c$.
Within the numerical treatment of this paper, the jump appears to be a first-order transition rather than a sharp crossover. 

\begin{figure}
\centerline{
\includegraphics[width=7.0cm]{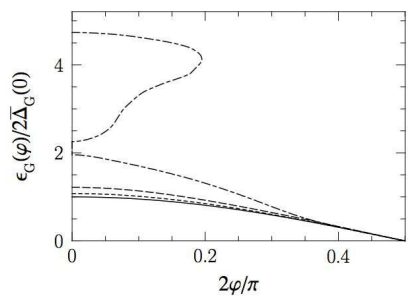}
}
\caption[1]{Anisotropy of the effective gap $\epsilon_{\rm G}(\varphi)$ for $\theta_J/\nu_1=4$ and $\gamma_1/\nu_1=0.1$:
(dotted line) $r_1=0.3$, (dashed line) $0.2$, and
(dot-dashed line) $0.19$.
For comparison, $\cos(2\varphi)$ is shown by a solid line.
Note that $\epsilon_{\rm G}(\varphi)$ is multi-valued for $r_1=0.19$; it is expected in such a case that the observed gap increases much more rapidly as $\varphi$ decreases than $\cos(2\varphi)$ does.
}
\label{fig_cosDep}
\end{figure}
\begin{figure}
\centerline{
\includegraphics[width=7.0cm]{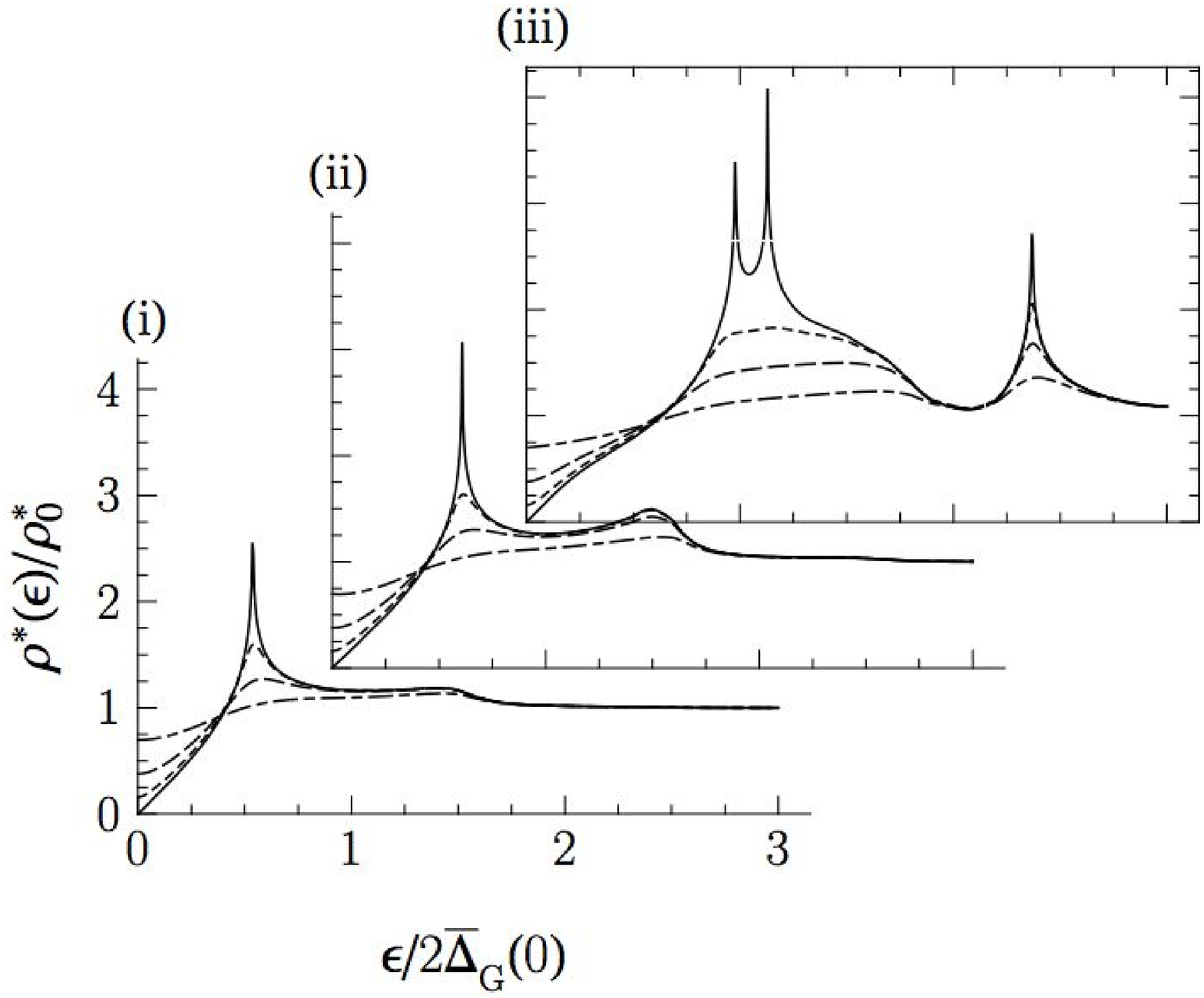}
}
\caption[1]{
Density of states $\rho^*(\epsilon)$ for $\theta_J/\nu_1=4$ and $\gamma_1/\nu_1=0.1$: (i) $r_1=0.3$, (ii) $0.2$, and (iii)  $019$; (solid line) $\gamma_Q/2\bar{\Delta}_{\rm G}(0)=0$, 
(dotted line) 0.03, (dashed line) 0.1, and (dot-dashed line) 0.3.
The sum rule for $\rho^*(\epsilon)$ that is significantly violated in the case of $r_1=0.19$ is approximately recovered when the phenomenological lifetime width $\gamma_Q$ is introduced.
}
\label{fig_rhoCrt}
\end{figure}

Since $\bar{\Delta}_{\rm G}(\epsilon+{\rm i}0)$ depends on $\epsilon$, the $\varphi$-dependence of the effective gap $\epsilon_{\rm G}(\varphi)$ defined by eq.~(\ref{EqEffectGap}) deviates from the $\cos(2\varphi)$ dependence, as shown in Fig.~\ref{fig_cosDep}.
The deviation of the gap anisotropy from the simple $d$-wave anisotropy has been actually observed in the cuprate oxide.\cite{Arpes-kondo}

Figure~\ref{fig_rhoCrt} shows the density of states when the lifetime width $\gamma_Q$ of Bogoliubov's quasi-particles is considered in eq.~(\ref{EqRho*}). The two low-energy peaks, which appear for $r_1=0.19$ and $\gamma_Q\rightarrow +0$, are not resolved even for small $\gamma_Q/2\bar{\Delta}_{\rm G}(0)=0.03$; the resolution is worse if the life-time effect is considered in the gap equation. It is likely that the two low-energy peaks are not resolved in the cuprate oxide even if the large-gap phase actually appears.

\mysection{Discussion}
\label{SecDiscussion} 
The exchange interaction $I_s^*(\omega+{\rm i}0,{\bm q})$ given by eq.~(\ref{EqIsStar1}) is composed of various terms:
\begin{align}
I_s^*(\omega+{\rm i}0,{\bm q}) &= 
I_s(\omega+{\rm i}0,{\bm q}) + \frac1{4}I_s^2(\omega+{\rm i}0,{\bm q})
\chi_s(\omega+{\rm i}0,{\bm q}),
\end{align}
where $I_s(\omega+{\rm i}0,{\bm q})$ is given by eq.~(\ref{EqIs3Terms}) or $I_s(\omega+{\rm i}0,{\bm q}) = J_s({\bm q}) +J_Q(\omega+{\rm i}0,{\bm q}) +J_\text{sf-ph}(\omega+{\rm i}0,{\bm q})$  and $\chi_s(\omega+{\rm i}0,{\bm q})$ is given by eq.~(\ref{EqDefIs}) or
$\chi_s(\omega+{\rm i}0,{\bm q})=\tilde{\chi}_s(\omega+{\rm i}0)/\left[1- (1/4)I_s(\omega+{\rm i}0,{\bm q})\tilde{\chi}_s(\omega+{\rm i}0)\right]$.
A crucial issue is to determine the main attractive interaction among the various terms that binds $d_{x^2-y^2}$-wave Cooper pairs.
According to Appendix~\ref{SecDynamicalSus}, ${\rm Im}\bigl[J_\text{sf-ph}(\omega+{\rm i}0,{\bm q})\bigr]$ has a sharp peak at approximately $\omega_{\rm ph}$ in the normal state and at approximately $\bar{\Delta}_{\rm G}(0) + \omega_{\rm ph}$ in the superconducting state;
when $J_\text{sf-ph}(\omega+{\rm i}0,{\bm q})$ has such a sharp peak, it is probable that $\chi_s(\omega+{\rm i}0,{\bm q})$ also has a similar structure in the same or a similar energy region.
According to the study in \mbox{\S\hskip2pt\ref{SecDipHump}},
the appearance of the dip-and-hump structures outside the coherence peaks can be explained by the existence of such a peak in the imaginary part of $I_s^*(\omega+{\rm i}0,{\bm q})$. If the dip-and-hump structure appears,
\begin{align}\label{EqJJ}
J_\text{sf-ph}(\omega+{\rm i}0,{\bm q}) 
+ \frac1{4}I_s^2(\omega+{\rm i}0,{\bm q})
\chi_s(\omega+{\rm i}0,{\bm q}) ,
\end{align}
must be at least as effective as 
the superexchange interaction $J_s({\bm q})$; if no dip-and hump structure or only a small one appears, the superexchange interaction $J_s({\bm q})$ must be the main attractive interaction.
It is certain that $\chi_s(\omega+{\rm i}0,{\bm q})$ is enhanced in the vicinity of the N\'{e}el state.
As studied in Appendix~\ref{SecDynamicalSus},
$J_\text{sf-ph}(\omega+{\rm i}0,{\bm q})$ includes the convolution of $\chi_s(\omega+{\rm i}0,{\bm q})$ and the Green function for phonons. 
Therefore, it is reasonable that the dip-and-hump structure is larger in under-doped cuprates than in over-doped cuprates.

The kink structure in the dispersion relation is mainly due to the on-site component of $I_s^*(\omega+{\rm i}0,{\bm q})$, and the dip-and-hump structure in the density of states is mainly due to the nearest-neighbor component of $I_s^*(\omega+{\rm i}0,{\bm q})$.
Although their appearances are closely related with each other, they may be different from each other; one may be observed but the other may not be observed, or their characteristic energy scales may be slightly different from each other such as $\nu_0\ne \nu_1$ and $\nu_0\simeq \nu_1$.

Three types of electron-phonon interaction are possible within the $t$-$J$ model:
one arising from the modulation of the band center or site energies by phonons,
one from that of transfer energies $t_{i\ne j}$ by phonons, and the one considered in this paper.
Since the first type of interaction couples with charge fluctuations, the single-site vertex correction divided by the mass enhancement factor is given by $\tilde{\phi}_c/\tilde{\phi}_{\rm e}$.
Since $\tilde{\phi}_c\ll 1$ and $\tilde{\phi}_{\rm e}\gg1$, the first type can never play any role in the $t$-$J$ model with almost half filling of electrons.
In the second type, no single-site vertex function can appear and $1/\tilde{\phi}_{\rm e}$ appears in every line of the electron Green functions. 
The second type cannot play a crucial role in the $t$-$J$ model.
Only the third type can play a crucial role in the $t$-$J$ model with almost half filling of electrons. 


If the electron-phonon interaction is ignored, the center wave number of antiferromagnetic spin fluctuations should be exactly ${\bm Q}_M=\left(\pm1,\pm1\right)(\pi/a)$ or very close to ${\bm Q}_M$.
When the electron-phonon interaction is strong, however, it cannot be ${\bm Q}_M$. 
The electron-phonon interaction can assist the development of antiferromagnetic spin fluctuations at wave numbers slightly different from ${\bm Q}_M$ and an antiferromagnetic order with such wave numbers.
It will be interesting to study by a microscopic and totally self-consistent theory whether or not the center wave numbers of coupled antiferromagnetic spin fluctuations are close to $\left(\pm 1, \pm 1 \pm 1/4\right)(\pi/a)$ and $\left(\pm 1\pm 1/4, \pm 1\right)(\pi/a)$ and whether those of coupled phonons are close to  $\left(0, \pm1/2\right)(\pi/a)$ and $\left(\pm 1/2, 0\right)(\pi/a)$; those of antiferromagnetic spin fluctuations have been observed in neutron-scattering experiments and those of phonons correspond to the so-called $4\times4$ checkerboard.
It has been proposed in a previous paper\cite{FJO-elph2} that the coexisting state of antiferromagnetic moments and lattice distortion with such wave numbers is simply the observed $4\times4$ checkerboard.
As studied in Appendix~\ref{SecSG-AM}, the coexistence of $d$-wave superconductivity and the antiferromagnetic state is possible, which implies that 
the coexistence of $d$-wave superconductivity and the checkerboard is also possible.

The so-called zero-temperature pseudogap is observed in under-doped cuprate superconductors.\cite{ZTPG1,ZTPG2,ZTPG3} A possible scenario to explain it is the coexistence of $d_{x^2-y^2}$-wave superconductivity and another order parameter, which may be conventional or exotic.
The large-gap phase studied in \mbox{\S\hskip2pt\ref{SecCrossover}} is another possible scenario that can explain the zero-temperature pseudogap.
The transition between the superconducting phase and the coexisting phase of superconductivity and the other order parameter must be a second-order  transition because the symmetries of the two phases are different from each other, while the transition between the large-gap and small-gap phases is a first-order transition or a sharp crossover because the symmetries of the two phase are the same as each other.
Nano-scale disorder or large inhomogeneity is observed in the gap structure of the pseudogap phase,\cite{ZTPG1,ZTPG2,ZTPG3} which appears to support the scenario of a first-order transition or a sharp crossover rather than that of a second-order transition, i.e., the large-gap phase rather than the coexisting phase.

Another type of pseudogap is observed in the critical region above $T_c$.
It is impossible to reproduce
the pseudogap by the theory developed in this paper, which is restricted to $T=0$~K.
In complete two dimensions, $T_c$ decreases to $+0$~K because of critical fluctuations.\cite{mermin} This fact implies that critical fluctuations are crucial in highly anisotropic quasi-two dimensions.
It has been proposed in previous papers\cite{FJP-PG1,FJP-PG2} that critical superconducting fluctuations are responsible for the pseudogap; the anisotropy of the pseudogap is such that its size is proportional to $\eta_{1d}^2({\bm k})$ or $\left[\cos(k_xa)-\cos(k_ya)\right]^2$, which is different from that of the superconducting gap. 
If the pseudogap opens, the
single-site susceptibility $\tilde{\chi}_s(0)$ is also enhanced even above $T_c$, as implied by results given in Appendix~\ref{SecSG-AM}.
If the enhancement of $\tilde{\chi}_s(0)$ is sufficiently large, antiferromagnetic moments can appear; thus, the coexistence of antiferromagnetic moments and lattice distortion, which is simply the checkerboard, can also appear due to the electron-phonon interaction.
The checkerboard has actually been observed above $T_c$. \cite{CheckerAboveTc}
It will be interesting to examine whether or not and how the appearance of the checkerboard and the development of the pseudogap are correlated.
Since the cuprate-oxide superconductor is in the vicinity of the N\'{e}el state, 
antiferromagnetic critical fluctuations are also so crucial that
they are responsible for the so-called $T$-linear resistivity.\cite{moriya}
It is straightforward to include the effects of both superconducting and antiferromagnetic critical fluctuations within the framework of this paper.

\mysection{Conclusion}
\label{SecConclusion}
In the vicinity of the Mott transition, a strong electron-phonon interaction arises from the modulation of the superexchange interaction by phonons.  The Kondo-lattice
theory of strong-coupling superconductivity based on the $t$-$J$ model with the electron-phonon interaction is formulated.
The self-energy of electrons is decomposed into the single-site and multisite self-energies. The calculation of the single-site self-energy, which should be self-consistently determined with other properties such as the superconducting order parameter, the multisite self-energy, the phonon self-energy, the total exchange interaction, and so forth, is reduced or mapped to a problem of self-consistently determining and  solving the Anderson model. It is proved that the ground state of the mapped Anderson model is a conventional Fermi liquid, i.e., the single-site self-energy is that of a conventional Fermi liquid even if the order parameter is nonzero or the multisite self-energy is anomalous.
The Fermi liquid characterized by the self-energy for the mapped Anderson model is further stabilized by the RVB mechanism.
The density of states for electrons near the chemical potential
is reduced by the RVB stabilization mechanism; it can be vanishingly small in a certain limiting case, although the density of states for quasi-particles, which is directly related to the specific heat coefficient, is still nonzero and can be large. 
The stabilized Fermi liquid is a relevant {\em unperturbed} state that can be used to study superconductivity and anomalous Fermi-liquid behaviors in the vicinity of the Mott transition or in the cuprate oxide.


Not only the superexchange interaction, which arises from the virtual exchange of a pair excitations of electrons between the upper and lower Hubbard bands, but also the exchange interactions arising from that of a pair excitation of quasi-particles and that of a coupled excitation of  spin fluctuations and phonons play a crucial role in the binding of $d_{x^2-y^2}$-wave Cooper pairs. 
On the basis of the analysis of the dynamical spin susceptibility, it is assumed in this paper that the imaginary part of the total exchange interaction has a sharp peak or dip at $\pm\omega^*$, where $\omega^*\simeq \omega_{\rm ph}$ in the normal state and $\frac1{2}\epsilon_{\rm G}\lesssim \omega^*\lesssim \frac1{2}\epsilon_{\rm G}+\omega_{\rm ph}$ in the superconducting state, where $\omega_{\rm ph}$ is the energy of relevant phonons and $\epsilon_{\rm G}$ is the superconducting gap.
Then, it was shown that the dispersion relation of quasi-particles has kink structures at approximately $\pm \omega^*$ above and below the chemical potential, the density of states for quasi-particles
has dip-and-hump structures at approximately $\pm \omega^*$ outside the coherence peaks  in the superconducting state, and
the anisotropy of the superconducting gap deviates from the simple anisotropy of $d_{x^2-y^2}$-wave superconductivity. 
These strong-coupling phenomena are consistent with observations in the cuprate-oxide superconductor.
As a functional of the exchange interaction or a function of the electron density, a first-order transition or a sharp crossover can occur  between small-gap and large-gap phases in the vicinity of antiferromagnetic instability.  It will be interesting to examine whether or not the large-gap phase is responsible for the zero-temperature pseudo-gap.

In this paper, numerous phenomenological parameters and simplifications were introduced to simplify the numerical processes.
It is desirable to develop a microscopic and totally self-consistent theory based on the formulation in this paper, in particular, to determine whether or not the imaginary parts of the spin susceptibility and the total exchange interaction actually have sharp peaks or dips at approximately $\pm\omega^*$, as assumed in this paper.

\appendix
\mysection{Proof of the Inequality of eq.~(\ref{EqSelfDelta})} 
\label{SecProof}
In this Appendix, it is only assumed that $\Sigma_{\sigma}(\epsilon \pm {\rm i}0,{\bm k})$'s are analytical in the upper and lower half planes, respectively; they may or may not be singular. 
The inverse of the diagonal Green function is given by
%
\begin{align}
G_{\sigma}^{-1}(\epsilon + {\rm i}0,{\bm k}) &=
\epsilon+\mu - E({\bm k})-\Sigma_{\sigma}(\epsilon + {\rm i}0,{\bm k}) 
\nonumber \\ & \quad
- \frac{|\Delta_{\sigma}(\epsilon + {\rm i}0,{\bm k})|^2}
{\epsilon-\mu + E(-{\bm k})+\Sigma_{-\sigma}(-\epsilon - {\rm i}0,-{\bm k})}.
\end{align}
We define the following real functions:
\begin{equation}
S_1(\epsilon,{\bm k}) =
\text{Re} G_{\sigma}^{-1}(\epsilon+{\rm i}0,{\bm k}),
\end{equation}
\begin{equation}
S_2(\epsilon,{\bm k})= 
\text{Im} G_{\sigma}^{-1}(\epsilon+{\rm i}0,{\bm k}),
\end{equation}
\begin{equation}
Y_n(\epsilon) = \frac1{N} \sum_{\bm k}
\frac{S_1^n(\epsilon,{\bm k} )}
{S_1^2(\epsilon,{\bm k}) +S_2^2(\epsilon,{\bm k})},
\end{equation}
\begin{equation}
Z_n(\epsilon) = \frac1{N} \sum_{\bm k}
\frac{S_2^n(\epsilon,{\bm k} )}
{S_1^2(\epsilon,{\bm k}) +S_2^2(\epsilon,{\bm k})},
\end{equation}
and
\begin{equation}
\tilde{S}_2(\epsilon) = 
-\mbox{Im} \bigl[ 
\Gamma (\epsilon + {\rm i}0)
+ \tilde{\Sigma}_\sigma(\epsilon + {\rm i}0) \bigr] .
\end{equation}
In general,
\begin{equation}\label{EqPositiveS2}
S_2(\epsilon,{\bm k}) \ge 
\tilde{S}_2(\epsilon) > 0,
\end{equation}
for any ${\bm k}$.
The single-site Green function is given by
%
$G_{ii\sigma}(\epsilon+{\rm i}0) =
Y_1(\epsilon) - i Z_1(\epsilon)$.
%
According to the mapping condition (\ref{EqMap-G1}), 
\begin{align}\label{EqDeltaGX}
\Delta(\epsilon) &=
%
%
-\text{Im} \Gamma (\epsilon \!+\! {\rm i}0) 
+\frac{X(\epsilon)}
{Y_1^2(\epsilon) + Z_1^2(\epsilon)}  ,
\end{align}
where
\begin{equation}
X(\epsilon) = Z_1(\epsilon)-\tilde{S}_2(\epsilon)[
Y_1^2(\epsilon) + Z_1^2(\epsilon)] .
\end{equation}
It is trivial that
$Y_0(\epsilon) = Z_0(\epsilon)$,
%
\begin{equation}\label{EqXYZ-1}
Y_2(\epsilon) + Z_2(\epsilon) = 1,
\end{equation}
and
\begin{equation}\label{EqXYZ-2}
Z_1(\epsilon) \ge
\tilde{S}_2(\epsilon) Y_0(\epsilon) =
\tilde{S}_2(\epsilon) Z_0(\epsilon) .
\end{equation}
According to eqs.~(\ref{EqXYZ-1}) and (\ref{EqXYZ-2}), it follows that
\begin{align}\label{EqXiLarger0}
X(\epsilon) &= 
Z_1(\epsilon)\left[Y_2(\epsilon) + Z_2(\epsilon)\right]
-\tilde{S}_2(\epsilon)[
Y_1^2(\epsilon) + Z_1^2(\epsilon)]
\nonumber \\ &\ge
\tilde{S}_2(\epsilon) \left[- Y_1^2(\epsilon) + Y_0(\epsilon)Y_2(\epsilon)\right]
\nonumber \\ & \quad 
+\tilde{S}_2(\epsilon) \left[ -Z_1^2(\epsilon) + Z_0(\epsilon)Z_2(\epsilon)
\right] .
\end{align}
Since the inequalities of
\begin{equation}
\frac1{N} \sum_{\bm k}
\frac{\left[x + S_1(\epsilon,{\bm k} )\right]^2}
{S_1^2(\epsilon,{\bm k}) +S_2^2(\epsilon,{\bm k})} >0,
\end{equation}
and
\begin{equation}
\frac1{N} \sum_{\bm k}
\frac{\left[x + S_2(\epsilon,{\bm k} )\right]^2}
{S_1^2(\epsilon,{\bm k}) +S_2^2(\epsilon,{\bm k})} >0,
\end{equation}
i.e., 
$Y_0(\epsilon) x^2 + 2 Y_1(\epsilon) x + Y_2(\epsilon) >0$
and 
$Z_0(\epsilon) x^2 + 2 Z_1(\epsilon) x + Z_2(\epsilon) >0$,
hold for any real $x$,
it follows that
\begin{equation}\label{EqY1P}
Y_1^2(\epsilon) - Y_0(\epsilon)Y_2(\epsilon) <0,
\end{equation}
and
\begin{equation}\label{EqZ1P}
Z_1^2(\epsilon) - Z_0(\epsilon)Z_2(\epsilon) <0.
\end{equation}
According to eqs.~(\ref{EqPositiveS2}), (\ref{EqXiLarger0}), (\ref{EqY1P}), and (\ref{EqZ1P}), it follows that $X(\epsilon)>0$.
Thus, the inequality (\ref{EqSelfDelta}),
$\Delta_{\rm A}(\epsilon) \ge
-\text{Im} \Gamma (\epsilon + {\rm i}0)$,
holds as a result of eq.~(\ref{EqDeltaGX}) even if the total self-energy $\Sigma_\sigma(\epsilon \pm {\rm i}0,{\bm k})$ is divergent or the order parameter $\Delta_{\sigma}(\epsilon \pm {\rm i}0,{\bm k})$ is nonzero. 

\mysection{Dynamical Polarization Function in Spin Channels}
\label{SecDynamicalSus}

\begin{figure*}
\centerline{
\begin{minipage}{8cm}
\includegraphics[width=6.0cm]{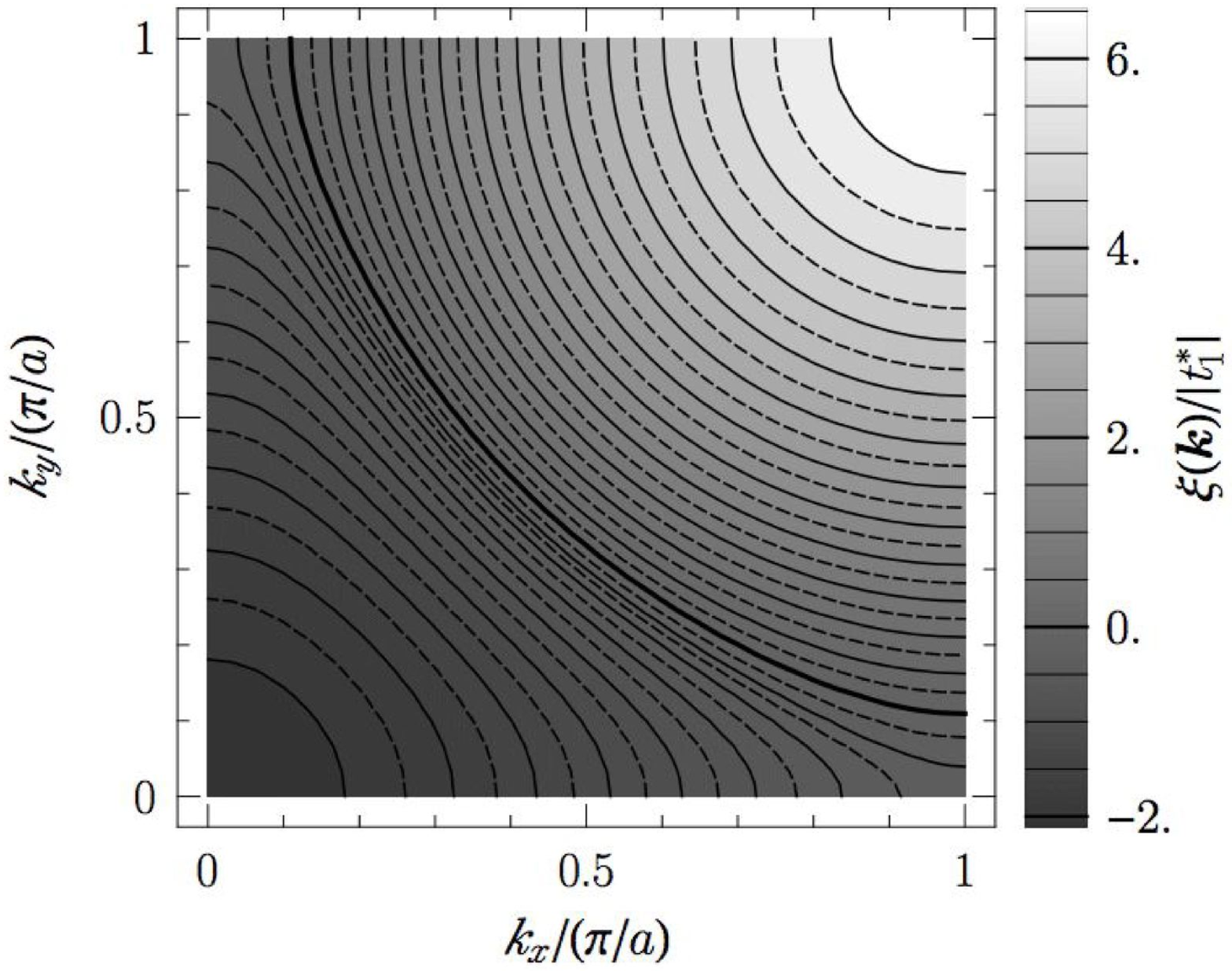}
\end{minipage}
\begin{minipage}{8cm}
\includegraphics[width=7.0cm]{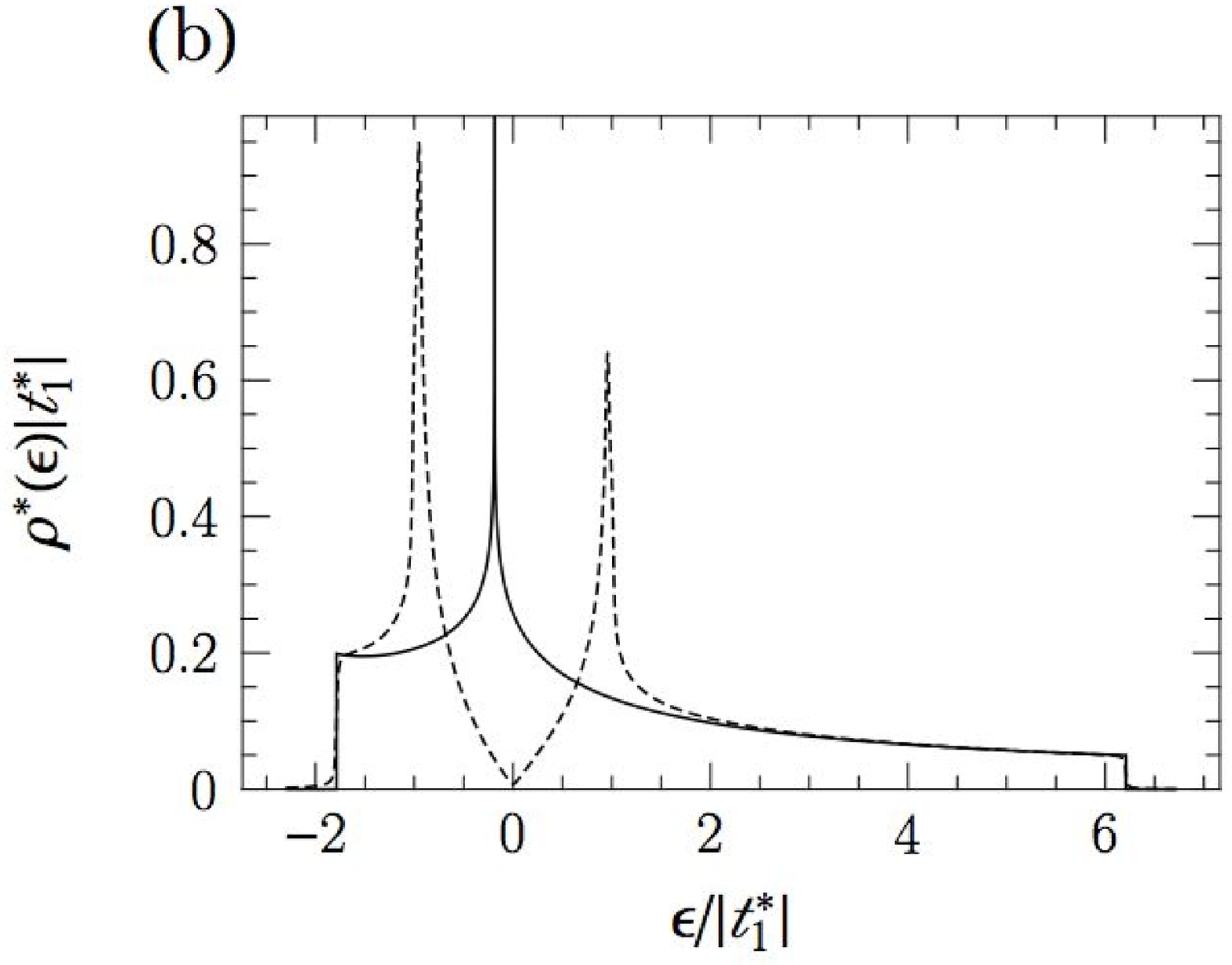}
\end{minipage}
}
\caption[1]{
(a) Fermi surfaces or lines in the normal state for the electron density  (dashed line) $n=0.1 i$ and (solid line) $n=0.1 i+0.05$, where $i$ is an integer, per unit cell. 
The thick solid line shows the Fermi surface for $n=0.85$.
(b) Density of states $\rho^*(\epsilon)$ for $n=0.85$: (solid line) the normal state and (dashed line) the superconducting state.
In the superconducting state, the logarithmic van Hove singularity is absorbed into the coherence peaks.
}
\label{fig_FSDOS}
\end{figure*}

It is assumed that the dispersion relation of quasi-particles is given by
\begin{align}
\xi({\bm k}) &= - 2t^*_1\left[\cos(k_xa)+\cos(k_ya)\right]
\nonumber \\ & \quad
- 4t^*_2 \cos(k_xa)\cos(k_ya)- \mu^* ,
\end{align}
where $t^*_1/|t^*_1|>0$, $t^*_2=-0.3t^*_1$, and $\mu^*$ is such that 
$(2/N)\sum_{\bm k}H\bigl[-\xi({\bm k})\bigr]
= 0.85$, i.e., $\mu^*=-1.0010 |t_1^*|$. According to the Fermi-surface sum rule, the electron number is 0.85 per unit cell in the normal state; it is also about 0.85 in the superconducting state. 
The Fermi surface or line for this set of parameters is similar to the observed one, as shown in Fig.~\ref{fig_FSDOS}(a). 
The Green function for quasi-particles is given by
\begin{align}
\bar{\cal G}_\sigma(\varepsilon\pm {\rm i}0,{\bm k})&=
\left(\begin{array}{cc}
\varepsilon - \xi({\bm k}) \pm {\rm i}\gamma_Q  & 
-\frac1{2}\eta_{1d}({\bm k})\bar{\Delta}_{\rm G}\\
-\frac1{2}\eta_{1d}({\bm k})\bar{\Delta}_{\rm G} &
 \varepsilon + \xi({\bm k}) \pm  {\rm i}\gamma_Q 
\end{array}\right)^{-1} ,
\end{align}
where $\eta_{1d}({\bm k}) = \cos(k_xa)-\cos(k_ya)$.
The energy dependence of the gap function $\bar{\Delta}_{\rm G}$ is ignored and the phase of $\bar{\Delta}_{\rm G}$ is chosen in such a way that
$\bar{\Delta}_{\rm G}$ is positive. A small phenomenological lifetime width $\gamma_Q=0.01|t^*_1|$ is assumed for convenience in numerical processes if necessary.
Figure~\ref{fig_FSDOS}(b) shows the density of states $\rho^*(\epsilon)$, which is defined by eq.~(\ref{EqRho*}), in the two cases of
$\bar{\Delta}_{\rm G}/|t^*_1|=0$ and $\bar{\Delta}_{\rm G}/|t^*_1|=1$, i.e., in the normal and superconducting states; the superconducting gap is as large as $2\bar{\Delta}_{\rm G}=2|t^*_1|$.

\begin{figure*}
\centerline{
\includegraphics[width=7.0cm]{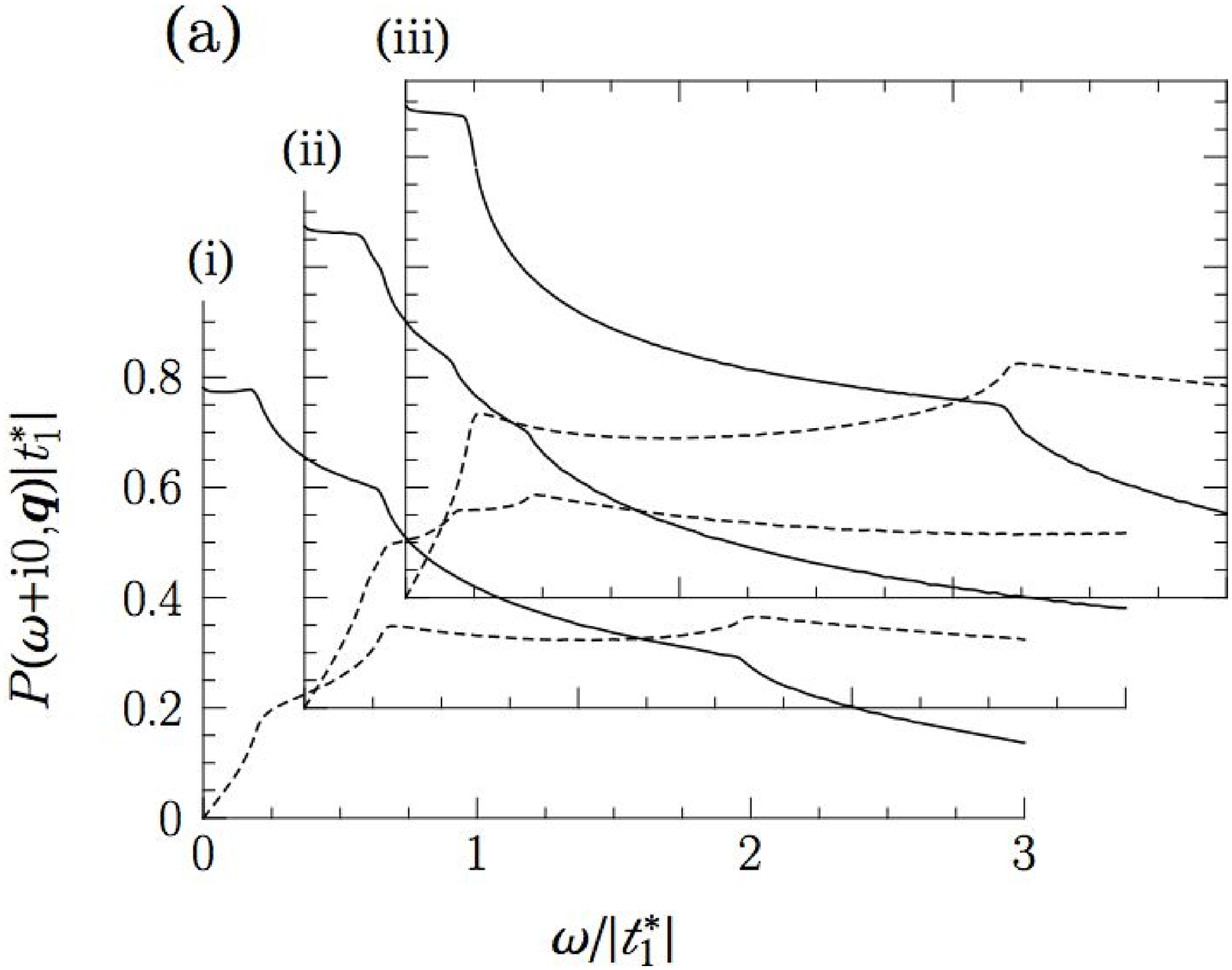}
\includegraphics[width=7.0cm]{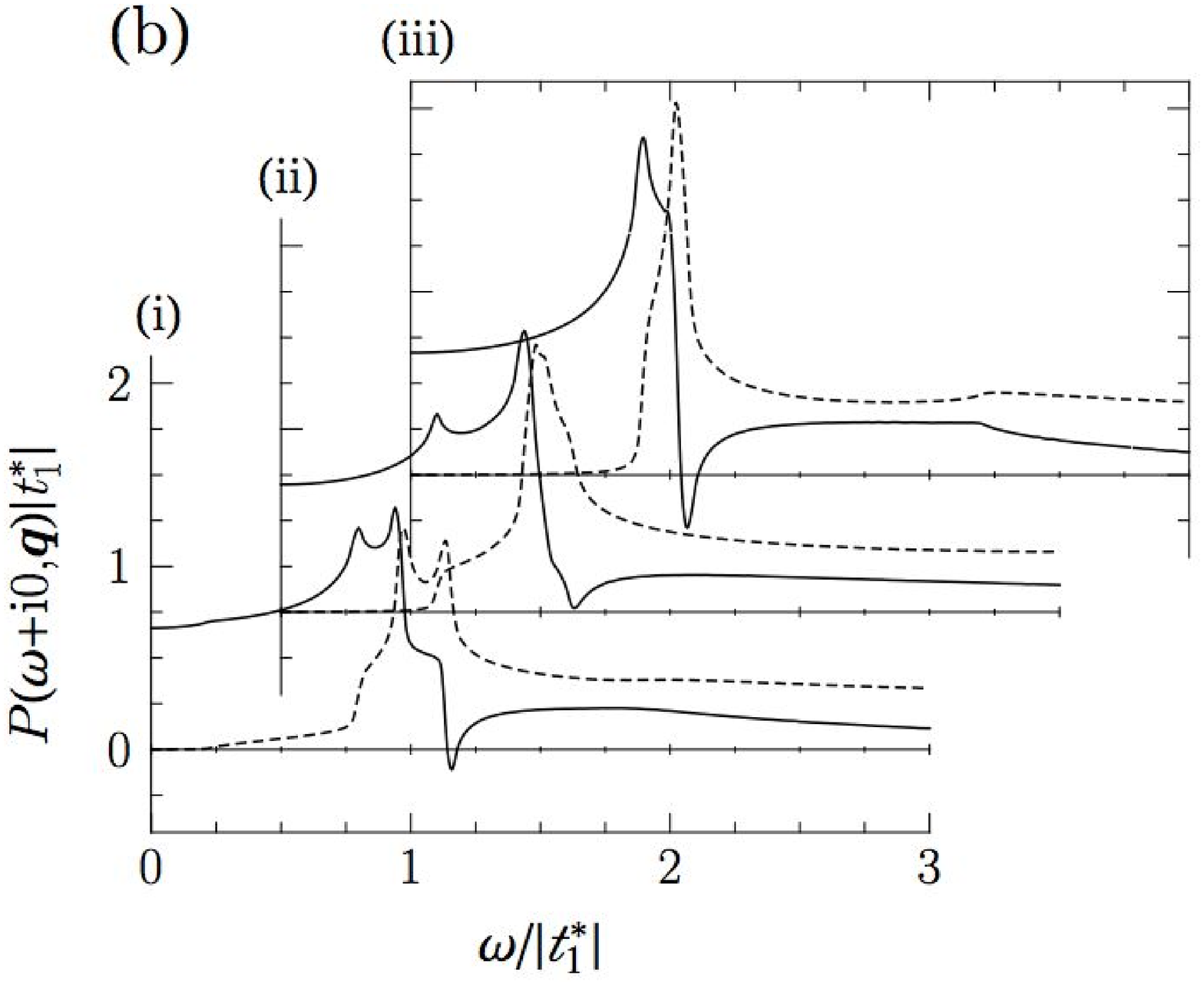}
}
\caption[1]{Polarization function $P(\omega+{\rm i}0,{\bm q})$: 
(a) the normal state and (b) the superconducting state with $|\bar{\Delta}_{\rm G}/t^*_1|=1$;  (i) ${\bm q}/(\pi/a)=(0.8,0.8)$, (ii) $(1,0.8)$, and (iii) $(1,1)$.
Solid and dotted lines show the real and imaginary parts, respectively.
}
\label{fig_Pi}
\end{figure*}
\begin{figure*}
\centerline{
\includegraphics[width=7.0cm]{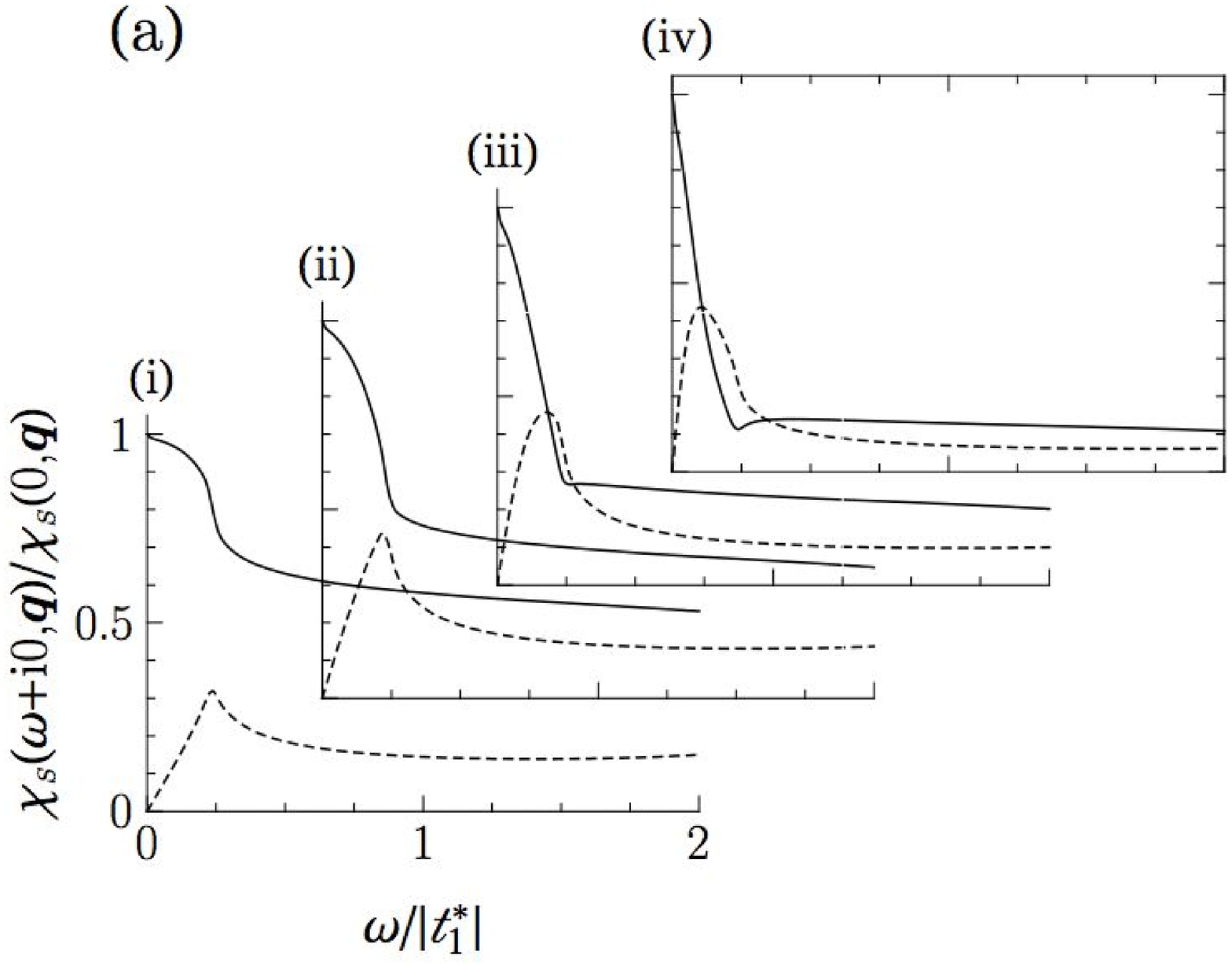}
\includegraphics[width=7.0cm]{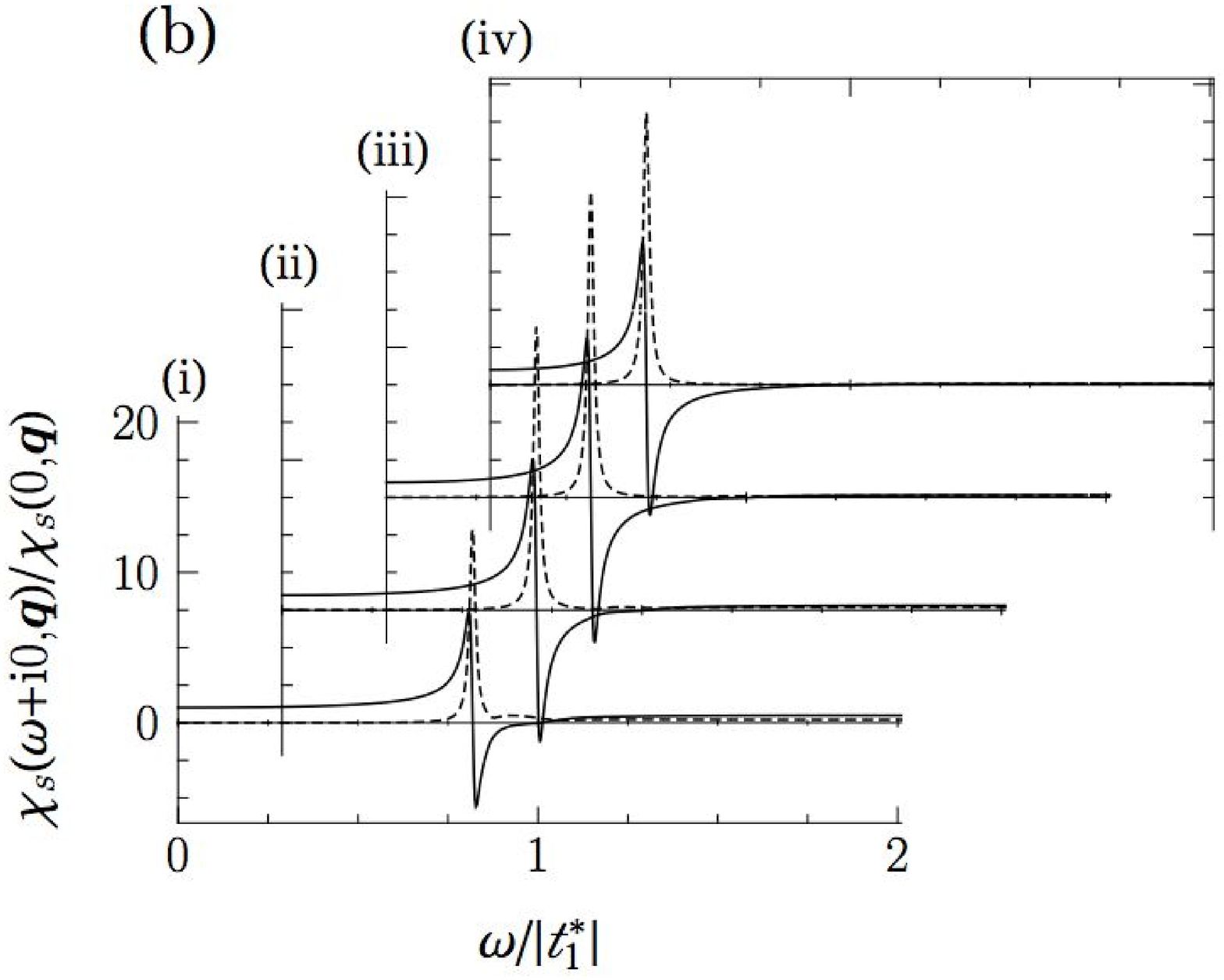}
}
\caption[1]{
$\chi_s(\omega+{\rm i}0,{\bm q})/\chi_s(0,{\bm q})$ for ${\bm q}=(\pm 1,\pm 1)(\pi/a)$:
(a) the normal state and (b) the superconducting state with $|\bar{\Delta}_{\rm G}/t^*_1|=1$; 
(i) $\kappa_{\bm q}=0.8$, (ii) $0.4$, (iii) $=0.2$, and (iv) $0.1$.
Solid and dashed lines show the real and imaginary parts, respectively.
}
\label{fig_chi}
\end{figure*}

Figure~\ref{fig_Pi} shows the polarization function $P(\omega+{\rm i}0,{\bm q})$, which is defined by eq.~(\ref{EqPolP}), in the normal and superconducting states for only three typical values of ${\bm q}$ near ${\bm Q}_M=(\pm 1, \pm 1)(\pi/a)$ among many calculated values of ${\bm q}$, which are distributed in the whole Brillouin zone.  
In the normal state, the real part of the static $P(0,{\bm q})$ has a broad peak near ${\bm Q}_M$ when it is plotted as a function of ${\bm q}$. Since the peak is broad, the electron-phonon interaction can play a crucial role in determining the center wave number ${\bm Q}_{\rm sf}$ of spin fluctuations or ${\bm Q}_{\rm AF}$ at which the static spin susceptibility $\chi(0,{\bm q})$ is maximum.
In the superconducting state, the imaginary part of $P(\omega+{\rm i}0,{\bm q})$ is suppressed in the low-energy region of $|\omega|\lesssim \bar{\Delta}_{\rm G}$ by the opening of the gap and it has a peak near $|\omega|\simeq \bar{\Delta}_{\rm G}$ .
The real part is enhanced in the low-energy region of $\omega\simeq\bar{\Delta}_{\rm G}$ by the opening of the gap unless $|{\bm q}| a\ll \pi$, which means that the opening of the superconducting gap assists the development of low-energy antiferromagnetic fluctuations and the appearance of antiferromagnetic moments.

When the electron-phonon interaction is ignored, the spin susceptibility is described by
\begin{align}
\frac1{\chi_s(\omega+{\rm i}0,{\bm q})} =
\frac{\tilde{W_s^2}}{\tilde{\chi}_s^2(0)}\left[
P_0(\omega+{\rm i}0,{\bm q})- P(\omega+{\rm i}0,{\bm q}) \right],
\end{align}
where
\begin{align}
P_0(\omega+{\rm i}0,{\bm q}) = 
\frac{\tilde{\chi}_s^2(0)}{\tilde{W}_s^2}
\left[
\frac1{\tilde{\chi}_s(\omega+{\rm i}0)} - J_s({\bm q})\right]
+ \tilde{P}(\omega+{\rm i}0) .
\end{align}
Following a previous study\cite{omega-linear} on the $\omega$-linear imaginary part of the susceptibility of the Anderson model,
it can be shown that 
\begin{align}
\lim_{\omega\rightarrow 0}{\rm Im}\frac1{\omega}\left[
\frac{\tilde{\chi}_s^2(0)}{\tilde{W_s^2}}\frac1{\tilde{\chi}_s(\omega+{\rm i}0)}- \tilde{P}(\omega+{\rm i}0)\right] =0.
\end{align}
Since the $\omega$-linear imaginary part vanishes, 
the energy dependence of $P_0(\omega+{\rm i}0,{\bm q})$ is ignored in the following part.
When a phenomenological parameter $\kappa_{\bm q}$ is defined by
$P_0(0,{\bm q}) = P(0,{\bm q})\left[1+\kappa_{\bm q}\right]$,
it follows that
\begin{align}
\frac{\chi_s(\omega+{\rm i}0,{\bm q})}{\chi_s(0,{\bm q})} =
\frac{\kappa_{\bm q}}{\displaystyle 
1+\kappa_{\bm q}
- \frac{P(\omega+{\rm i}0,{\bm q})}{P(0,{\bm q})}}.
\end{align}
Figure~\ref{fig_chi} shows $\chi_s(\omega+{\rm i}0,{\bm q})/\chi_s(0,{\bm q})$ for ${\bm q}={\bm Q}_M$ and four values of $\kappa_{\bm q}$ in the normal and superconducting states.
Because of the suppression of the imaginary part of $P(\omega+{\rm i}0,{\bm q})$ in the superconducting state, the imaginary part of $\chi(\omega+{\rm i}0.{\bm q})$ shows a sharp peak near $\omega\simeq \bar{\Delta}_{\rm G}$ and a sharp dip near $\omega\simeq - \bar{\Delta}_{\rm G}$.

The  susceptibility and the Green function for phonons are described in the spectral representation:
\begin{align} 
\chi_s({\rm i}\omega_\ell,{\bm q})=\int_{0}^{+\infty} d\epsilon
A_{\text{s}{\bm q}}(\epsilon)\left[\frac1{{\rm i}\omega_\ell - \epsilon}
- \frac1{{\rm i}\omega_\ell + \epsilon} \right],
\end{align}
and
\begin{align} 
G_\lambda({\rm i}\omega_\ell,{\bm q})=\int_{0}^{+\infty} d\epsilon
A_{\lambda{\bm q}}(\epsilon)\left[\frac1{{\rm i}\omega_\ell - \epsilon}
- \frac1{{\rm i}\omega_\ell + \epsilon} \right].
\end{align}
At $T=0$~K, it follows that
\begin{align}\label{EqConvolution}
k_{\rm B}T & \sum_{\omega_n}G_\lambda({\rm i}\omega_\ell+{\rm i}\omega_n,{\bm q})
\chi_s({\rm i}\omega_n,{\bm p})
\nonumber \\ & =
\int_{0}^{+\infty} \hskip-15pt d\epsilon A_{{\bm q},{\bm p}}(\epsilon)
\left[\frac1{{\rm i}\omega_\ell - \epsilon}
- \frac1{{\rm i}\omega_\ell + \epsilon} \right],
\end{align}
where
\begin{align}
A_{{\bm q},{\bm p}}(\epsilon) = 
\int_{0}^{\epsilon} d\epsilon^\prime
A_{\lambda{\bm q}}(\epsilon-\epsilon^\prime)A_{\text{s}{\bm p}}(\epsilon^\prime).
\end{align}
Note that eq.~(\ref{EqConvolution}) appears in $J_\text{sf-ph}({\rm i}\omega_\ell,{\bm q})$ defined by eq.~(\ref{EqJsf-ph}).
As is shown in Fig.~\ref{fig_chi}(b), $A_{\text{s}{\bm q}}(\omega)$ with ${\bm q}={\bm Q}_M$ has sharp peaks near $\omega =\bar{\Delta}_{\rm G}$ in the superconducting state.
Provided that $A_{\lambda{\bm q}}(\omega)$ has a sharp peak near $\omega=\omega_{\rm ph}$, 
$A_{{\bm q},{\bm p}}(\epsilon)$ also has a sharp peak near $\omega=\bar{\Delta}_{\rm G} + \omega_\text{ph}$, so that
$J_\text{sf-ph}(\omega+{\rm i}0,{\bm q})$ has
a sharp peak near $\omega=\bar{\Delta}_{\rm G} + \omega_\text{ph}$.

In the complete theory, both $\chi_s({\rm i}\omega_\ell,{\bm q})$ and
$G_\lambda({\rm i}\omega_\ell,{\bm q})$ should be self-consistently calculated with many other quantities.
However, it is reasonable to expect that, in the final self-consistent solution, ${\rm Im}\chi_s(\omega + {\rm i}0,{\bm q}\simeq {\bm Q}_M)$ will have a sharp peak at $\omega\simeq \omega_{\rm ph}$ in the normal state and
at $\omega$ such that $\bar{\Delta}_{\rm G} \lesssim \omega \lesssim \bar{\Delta}_{\rm G} + \omega_{\rm ph}$ in the superconducting state.


\mysection{Coexistence of Superconductivity and Antiferromagnetism}
\label{SecSG-AM}
According to eq.~(\ref{EqRhoSSA}),  it follows that
\begin{align}
\rho(\epsilon) = \frac{\Delta_{\rm A}(\epsilon)}
{\bigl\{{\rm Re}\bigl[1/\tilde{G}_\sigma(\epsilon+{\rm i}0)\bigr]\bigr\}^2 +\Delta_{\rm A}^2(\epsilon)   }.
\end{align}
In general, $\mbox{Re}[1/\tilde{G}_{\sigma}(+{\rm i}0)]\ne 0$ for the non-half filling of electrons $(n\ne 1)$. When a gap opens in $\rho(\epsilon)$, a gap also opens in $\Delta_{\rm A} (\varepsilon)$.   In this Appendix, the enhancement of the single-site spin susceptibility $\tilde{\chi}_s(0)$ by the opening of the gap is studied.

Although $\Delta_{\rm A}(\epsilon)$ should be self-consistently determined to satisfy the mapping condition (\ref{EqMap-G1}), the gap structure of $\Delta_{\rm A}(\varepsilon)$ is phenomenologically treated here.
Since $\rho(\epsilon) \propto \Delta_{\rm A}(\epsilon) \propto |\epsilon|$ for small $|\epsilon|$ at $T=0$~K for $d$-wave superconductivity,
\begin{equation}\label{EqPhDelta}
\Delta_{\rm A}(\epsilon) = \left\{\begin{array}{cc}
\Delta_1, &- D \le \epsilon  \le - \epsilon_0 \vspace{0.1cm}\\
\displaystyle \Delta_0 + 
\left(\Delta_1-\Delta_0\right) \left|\frac{\epsilon}{\epsilon_0}\right|, 
& -\epsilon_0 \le \epsilon \le 0 \vspace{0.15cm}\\
\displaystyle \Delta_0 + 
\left(\Delta_2-\Delta_0\right)\left|\frac{\epsilon}{\epsilon_0}\right|, 
& 0 \le \epsilon \le \epsilon_0 \vspace{0.1cm}\\
\Delta_2, &\epsilon_0 \le \epsilon \le  D \vspace{0.1cm}\\
0 ,& |\epsilon| >D 
\end{array}\right. ,
\end{equation}
where $\epsilon_0 \simeq \bar{\Delta}_{\rm G}$ and $D\simeq W/2$, are assumed. 
According to eq.~(\ref{EqMap-G1}),
$\Delta_1 \simeq O(W/\pi)$ and $\Delta_1 \gg \Delta_2$
because the chemical potential is at the top of the lower Hubbard band.
Although $\Delta_0=0$ at $T=0$~K, 
$\Delta_0$ is treated as another parameter.
It is assumed that $\Delta_1 \ge \Delta_0 \ge 0$.

Since $U_\infty/D=+\infty$, a doubly occupied configuration of $d$ electrons is not allowed in any eigenstate.
Thus, the lowest singlet state of the Anderson model is expanded in such a way that 
\begin{align}\label{EqPhi}
\Phi_{s} &= \Bigl[ A_0
+ \sum_{{\bm k}\sigma} 
A_{d;{\bm k}\sigma} d_{\sigma}^\dag c_{{\bm k}\sigma}
\nonumber \\ & \quad 
+\sum_{{\bm k}\sigma} \sum_{{\bm p}-\sigma}
A_{{\bm k}\sigma;{\bm p}-\sigma} c_{{\bm k}\sigma}^\dag
c_{{\bm p}-\sigma} + \cdots
\Bigr] \left|0\right> , \quad
\end{align}
where $\left|0\right>$ is the Fermi vacuum for conduction electrons with no $d$ electron.   The ground-state wave function $\Phi_{s}$ satisfies
\begin{equation}
\bigl({\cal H}_{\rm A}
-\tilde{\mu} {\cal N}_{\rm A} + {\cal H}_{\rm ext} \bigr)\Phi_{s} = E_s \Phi_{s},
\end{equation}
where ${\cal H}_{\rm A}$ is defined by eq.~(\ref{EqAnderson}), 
${\cal N}_{\rm A} = \sum_{\sigma}n_{d\sigma} +
\sum_{{\bm k}\sigma}c_{{\bm k}\sigma}^\dag c_{{\bm k}\sigma}$,
%
${\cal H}_{\rm ext} = - \sum_{\sigma}\left(
\Delta \tilde{\mu} +\sigma \tilde{h}\right)n_{d\sigma} $
are infinitesimally small external fields, and
$E_s$ is the energy of the singlet. 
When only $A_0$, $A_{d;{\bm k}\sigma}$, and 
$A_{{\bm k}\sigma;{\bm p}-\sigma}$ are considered, 
it follows that
\begin{equation}\label{EqGroundEs}
E_s = -  \frac{2}{\pi} \hskip-2pt
\int_{-D}^{0} \hskip-5pt d\epsilon \Delta_{\rm A}(\epsilon)
\frac{E_d- \Delta\tilde{\mu} - \epsilon - E_s  }{
\left(E_d- \Delta\tilde{\mu} - \epsilon -E_s  \right)^2 
- \tilde{h}^2 } ,
\end{equation}
where $E_d$ is the energy of the lowest doublet; 
\begin{equation}
E_d = E_0 +\epsilon_d -\tilde{\mu}
-\frac{\Delta_2}{\pi} \ln \frac{D}{\Delta_1},
\end{equation}
where $E_0$ is the energy of the Fermi vacuum.

It is easy to confirm that eq.~(\ref{EqGroundEs}) gives $E_s$ such that
$E_s<E_d$. When $\rho(\epsilon) \propto |\epsilon|$ for small $|\epsilon|$, the ground state of the Anderson model is a singlet; thus,  
the ground state of the $t$-$J$ model within the restricted Hilbert subspace where no order parameter exists is also a singlet, even if the reservoir effect is ignored. 
The $d$ electron density, which is simply denoted by $n$ here, is given by 
%
$n = - \partial E_s/\partial \Delta\mu $,
so that
\begin{equation}\label{EqSingleBinding}
\frac{n}{1-n} = 
\frac{2}{\pi}\int_{-D}^{0} 
d\epsilon  
\frac{\Delta_{\rm A} (\epsilon)}
{(E_d - \epsilon - E_s)^2 }.
\end{equation}
The magnetization is given by
$m = - \partial E_s/\partial h$
and the susceptibility is given by
$\tilde{\chi}_s(0) =
\bigl[m/\tilde{h}\bigr]_{\tilde{h}\rightarrow0}$ or
\begin{align}
\tilde{\chi}_s(0) &=
\frac{4(1-n)}{\pi}
\int_{-D}^{0} d\epsilon 
\frac{\Delta_{\rm A} (\epsilon)}
{(E_d - \epsilon  - E_s)^3}.\qquad 
\end{align}

\begin{figure}
\centerline{
\includegraphics[width=7.0cm]{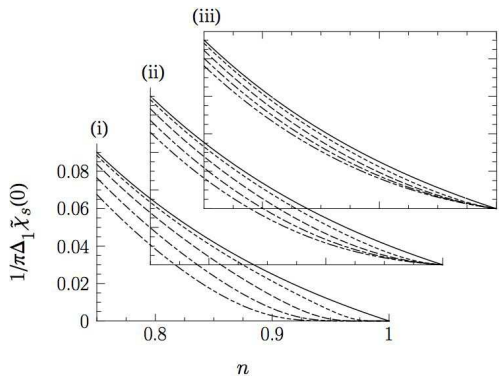}
}
\caption[1]{
$1/\pi \Delta_1 \tilde{\chi}_s(0)$ as a function of $n$:
(i) $\Delta_0/\Delta_1=0$, (ii) $\Delta_0/\Delta_1=0.1$, 
and (iii) $\Delta_0/\Delta_1=0.2$.
Solid, dotted, dashed, dot-dashed, and double-dot-dashed lines represent
$\varepsilon_0/\Delta_1=0$, 0.1, 0.2, 0.3, and 0.4, respectively;
every solid line shows the relation of eq.~(\ref{EqChiS0}).
}
\label{fig_chis}
\end{figure}

When
no gap structure develops in $\Delta_{\rm A}(\varepsilon)$,
i.e., $\Delta_0=\Delta_1$ or $\epsilon_0=0$, it follows that
$E_s = E_d - (2 \Delta_1/\pi)(1-n)/n$
and
\begin{equation}\label{EqChiS0}
\tilde{\chi}_s(0) =  \frac{\pi}{2\Delta_1}\frac{n^2}{1-n}.
\end{equation}
Since $D \gg \left|E_d-E_s\right|$, $D\rightarrow +\infty$ is assumed here.
When the gap develops in $\Delta_{\rm A}(\varepsilon)$, 
$\tilde{\chi}_s(0)$ is enhanced or 
$1/\tilde{\chi}_s(0)$ is reduced, as shown in Fig.~\ref{fig_chis}. 



The N\'{e}el temperature $T_{\rm N}$ of the $t$-$J$ model is defined by 
\begin{equation}\label{EqAF-Instability}
\left[\frac1{\tilde{\chi}_s(0)} - \frac1{4}J_s({\bm q})
- \frac1{4} J_Q(0,{\bm q}) -\frac{1}{4} J_\text{sf-ph} (0,{\bm q}) \right]_{T=T_{\rm N}}\!=0,
\end{equation}
where $T_{\rm N}$ should be maximized as a function of ${\bm q}$. 
As shown in Fig.~\ref{fig_chis}, $1/\tilde{\chi}_s(0)$ is greatly reduced  when a large gap opens, for example, when $\Delta_0/\Delta_1 \simeq 0$, $\varepsilon_0/\Delta_1\gtrsim 0.2$, and $n\gtrsim0.9$.
On the other hand, the superexchange interaction $J_s({\bm q})$ is not reduced by the opening of the gap.
As shown in Fig.~\ref{fig_Pi}, $P(0,{\bm q})$ is not reduced at least unless $|{\bm q}|a \ll \pi$.
According to eq.~(\ref{EqJQ}), $P(0,{\bm q})$ is the main term of $J_Q(0,{\bm q})$. Thus, $J_Q(0,{\bm q})$
is not reduced unless $|{\bm q}|a \ll \pi$, which implies that $J_\text{sf-ph} (0,{\bm q})$ is also not reduced unless $|{\bm q}|a \ll \pi$.
If a superconducting gap $2\bar{\Delta}_{\rm G}$ as large as $2\bar{\Delta}_{\rm G}/\Delta_1 \gtrsim 0.4$ opens, it is probable that antiferromagnetism appears for $n\gtrsim 0.9$, at least at $T=0$~K, i.e., antiferromagnetism coexists with superconductivity for $n \gtrsim0.9$.




\begin{thebibliography}{99}
\bibitem{comSquareLattice}
The notation $M$ for the square lattice is used in this paper.
\bibitem{comFilling}
The case of more than half filling, $1<n<2$, can be treated by making use of the particle-hole transformation.
\end{thebibliography}
\end{document}